\documentclass[11pt, a4paper, logo, copyright]{fdtn_template/fdtn}

\usepackage[numbers, sort]{natbib}
\usepackage{caption}
\usepackage{xspace}
\usepackage[table]{xcolor}

\usepackage{tikz}
\usepackage{pgfplots}
\pgfplotsset{compat=1.18}
\usetikzlibrary{positioning,arrows.meta,fit,calc,backgrounds}
\usepackage{multirow}
\usepackage{hyperref}
\usepackage{amsfonts}
\usepackage{amsmath}
\usepackage{amssymb}
\usepackage{fontawesome5}

\usepackage[bottom]{footmisc}
\usepackage{setspace}

\usepackage{graphicx}
\usepackage{array, booktabs, tabularx}
\hypersetup{hidelinks=true, colorlinks=true, citecolor=blue, linkcolor=black, urlcolor=black}

\usepackage{inconsolata}

\usepackage[utf8]{inputenc}
\usepackage[T1]{fontenc}
\usepackage{url}
\usepackage{nicefrac}
\usepackage{microtype}

\usepackage{tcolorbox}
\tcbuselibrary{breakable}

\newcommand{\nummodels}{27}

\newcommand{\opencode}[1][1em]{%
  \raisebox{-0.2\height}{\includegraphics[height=#1]{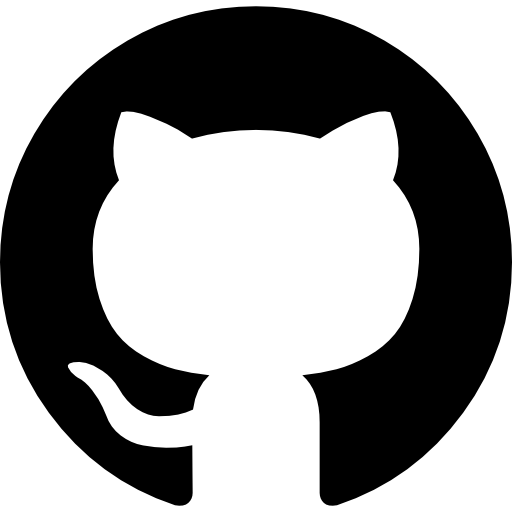}}%
}

\addto\extrasenglish{
}

\reportnumber{002}

\renewcommand{\phi}{\varphi}

\renewcommand{\leq}{\leqslant}
\renewcommand{\geq}{\geqslant}

\renewcommand{\epsilon}{\varepsilon}
\renewcommand{\imath}{\mathrm{i}}

\newlength{\restsubwidth}
\newlength{\restsubheight}
\newlength{\restsubmoreheight}
\newcommand{\rest}[2]{%
        \settowidth{\restsubwidth}{\ensuremath{#2}}
        \settoheight{\restsubheight}{\ensuremath{{}_{#2}}}
        \ensuremath{{#1\hskip 0.5pt}_{\vrule\kern2pt\parbox[b][%
        4pt][b]{\the\restsubwidth}{%
                        \ensuremath{{}_{#2}}}}}
        }

\definecolor{customlink}{HTML}{A9336A}
\renewcommand{\today}{}

\title{Vulnerability Localization Benchmark: Measuring Agentic Security Analysis at Repository Scale}

\author{
    Aman Priyanshu$^{*,1}$, 
    Supriti Vijay$^{*,1}$, 
    Kimia Majd$^{*, 1}$, 
    Xuhong He$^{1, 3, \dagger}$,
    Fraser Burch$^{1}$, 
    Takahiro Matsumoto$^{1}$, 
    Jianliang He$^{1, 2, \dagger}$, 
    Baturay Saglam$^{1, 2, \dagger}$, 
    Arthur Goldblatt$^{1}$, 
    Zhuoran Yang$^{1, 2, \dagger}$, 
    Amin Karbasi$^{1}$ \\
    {\normalsize $^{1}$Foundation AI--Cisco Systems Inc.} \\
    {\normalsize $^{2}$Yale University} \\
    {\normalsize $^{3}$Carnegie Mellon University} \\
    \vspace{1.0em}
    {\small $^{\dagger}$Work done while at Foundation AI} \\
    {\small $^{*}$Equal Contribution. Corresponding authors:
    \texttt{\{ampriyan,suprivij,kimia\}@cisco.com}}

\vskip 0.1in
\begin{center}
\small
\opencode~\href{https://github.com/cisco-foundation-ai/vulnerability-localization-benchmark}{Code \& Data}\quad
{\textcolor{red!50!black}{\faGlobe}}~\href{https://cisco-foundation-ai.github.io/vulnerability-localization-benchmark/}{Website}
\end{center}
\vskip -0.15in
}

\begin{document}

\begin{abstract}
Language-model agents increasingly operate over complete software repositories, yet cybersecurity evaluations primarily measure whether they can detect, reproduce, or repair vulnerabilities rather than whether they can locate the relevant code. We study \emph{vulnerability localization}: given a weakness class and an unfamiliar repository, identify the implementation files associated with that weakness. We introduce the Vulnerability Localization Benchmark (VLoc Bench), comprising 500 real world vulnerabilities from 290 repositories across six package ecosystems and 147 CWE categories. Each task pairs repository snapshots immediately before and after a security fix. On the vulnerable snapshot, an agent receives only the CWE description and read-only terminal access and must return the affected files; on the patched snapshot, it must determine that the recorded vulnerability is no longer present. We evaluate 27 language models and four static-analysis tools under a common agent interface. Repository-scale vulnerability localization remains difficult: the strongest system achieves 0.229 File F1, and 38.4\% of tasks receive no correct localization from any evaluated model. 

We further find that stronger localization does not imply reliable behavior after remediation: systems that identify vulnerable files effectively can still report unsupported locations on patched repositories. These results establish vulnerability localization as a distinct repository-scale capability and provide a setting for studying both how security agents search for vulnerable code and when they should refrain from reporting it.
\end{abstract}
\maketitle

\section{Introduction}
\label{sec:introduction}

Language model agents have become capable of software-engineering tasks that require reasoning over and modifying real codebases, and these capabilities increasingly extend to security analysis. A growing body of benchmarks evaluates agents on real vulnerabilities by asking them to generate a patch, produce an input that triggers vulnerable behavior, or determine whether a given piece of code is vulnerable~\cite{bui2026vul4py,wang2026cybergym,ding2025primevul}. These tasks measure important security capabilities, but they do not necessarily measure \emph{localization}: identifying where in a repository the vulnerable implementation lies. Detection and localization are distinct problems. Vulnerability detection asks whether code that has already been selected is vulnerable; vulnerability localization asks which code should have been selected in the first place. For a security engineer, this distinction is consequential: remediation begins by narrowing a repository to the files that require investigation, which determines what must be patched, what other components may be affected, and whether the same defect propagates to forks or vendored copies.

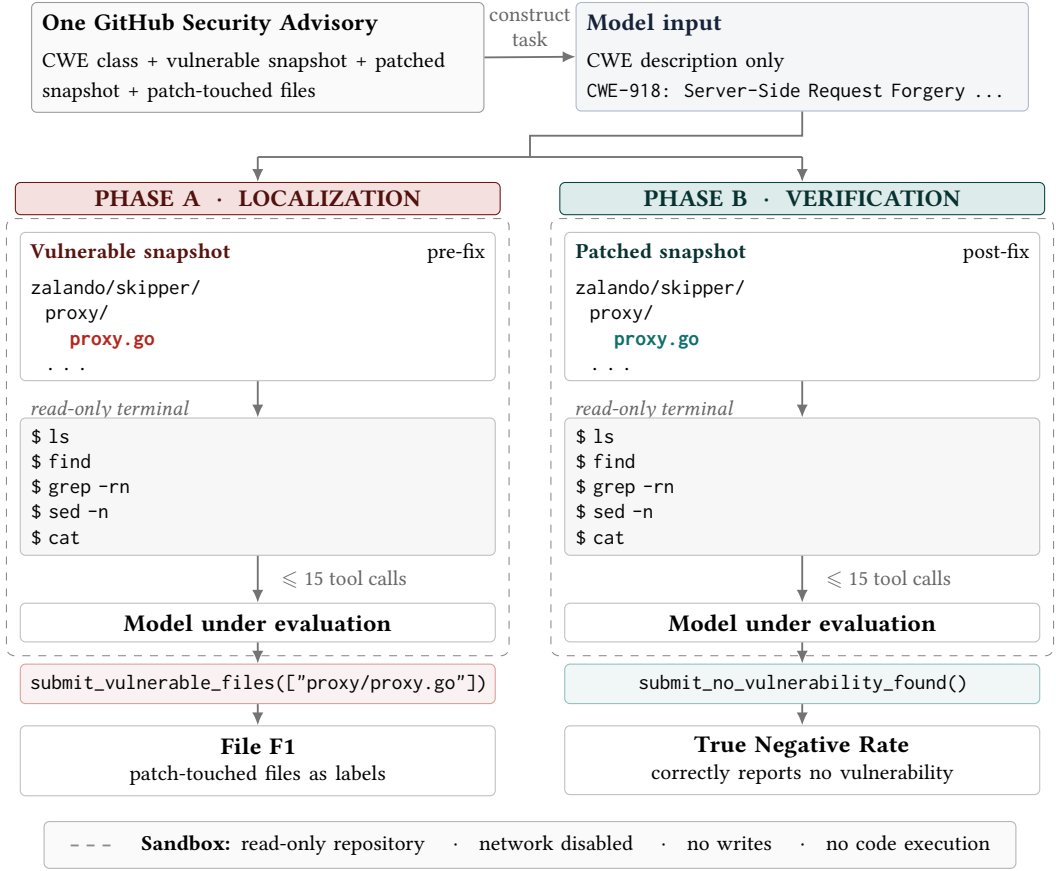
\begin{figure}[t]
\centering
\begin{tikzpicture}[
  font=\footnotesize,
  card/.style={
    rounded corners=2pt,
    draw=black!25,
    fill=white,
    inner sep=4pt,
    align=left,
    text width=6.0cm
  },
  repo/.style={
    rounded corners=2pt,
    draw=black!25,
    fill=white,
    inner sep=4pt,
    align=left,
    text width=6.0cm,
    font=\scriptsize
  },
  term/.style={
    rounded corners=2pt,
    draw=black!25,
    fill=black!3,
    inner sep=4pt,
    align=left,
    text width=6.0cm,
    font=\scriptsize
  },
  hdr/.style={
    rounded corners=2pt,
    draw=#1!60!black,
    fill=#1!14,
    inner sep=3pt,
    align=center,
    text width=6.2cm,
    font=\footnotesize\bfseries,
    text=#1!45!black
  },
  arr/.style={
    -{Latex[length=1.7mm,width=1.5mm]},
    draw=black!55,
    line width=0.7pt
  },
  note/.style={
    font=\scriptsize,
    text=black!60
  },
]

\definecolor{figvuln}{HTML}{B3261E}
\definecolor{figclean}{HTML}{0F6F6C}
\definecolor{figagent}{HTML}{2F4B7C}

% ============================================================
% Advisory -> model input
% ============================================================

\node[
  card,
  text width=5.7cm,
  minimum height=1.35cm,
  anchor=north,
  fill=black!2,
  draw=black!40
] (task) at (-3.6,0) {%
  \textbf{One GitHub Security Advisory}\\[3pt]
  {\scriptsize
  CWE class + vulnerable snapshot + patched snapshot +
  patch-touched files}
};

\node[
  card,
  text width=5.7cm,
  minimum height=1.35cm,
  anchor=north,
  fill=figagent!5,
  draw=figagent!40
] (given) at (3.6,0) {%
  \textcolor{figagent!45!black}{\textbf{Model input}}\\[3pt]
  {\scriptsize CWE description only}\\
  {\scriptsize\ttfamily
  CWE-918: Server-Side Request Forgery ...}
};

\draw[arr]
  (task.east) --
  node[note, above, align=center] {construct\\task}
  (given.west);

% ============================================================
% Fan-out
% ============================================================

\coordinate (split) at (0,-2.05);
\coordinate (busA) at (-3.6,-2.05);
\coordinate (busB) at ( 3.6,-2.05);

\draw[
  draw=black!55,
  line width=0.7pt
]
  (given.south) -- ++(0,-0.32) -| (split);

\draw[
  draw=black!55,
  line width=0.7pt
]
  (busA) -- (busB);

% ============================================================
% Phase headers
% ============================================================

\node[
  hdr=figvuln,
  anchor=north
] (hA) at (-3.6,-2.38)
  {PHASE A\ \ \textperiodcentered\ \ LOCALIZATION};

\node[
  hdr=figclean,
  anchor=north
] (hB) at (3.6,-2.38)
  {PHASE B\ \ \textperiodcentered\ \ VERIFICATION};

\draw[arr] (busA) -- (hA.north);
\draw[arr] (busB) -- (hB.north);

% ============================================================
% Phase A
% ============================================================

\node[
  repo,
  anchor=north,
  below=0.22cm of hA
] (snapA) {%
  \textbf{\textcolor{figvuln!50!black}{Vulnerable snapshot}}
  \hfill pre-fix\\[4pt]
  {\ttfamily
  zalando/skipper/\\
  \hspace*{1.5ex}proxy/\\
  \hspace*{4ex}\textcolor{figvuln}{\textbf{proxy.go}}\\
  \hspace*{1.5ex}\ldots}
};

\node[
  note,
  anchor=north west,
  font=\scriptsize\itshape
] (roA) at ($(snapA.south west)+(0,-0.12)$)
  {read-only terminal};

\node[
  term,
  anchor=north,
  below=0.48cm of snapA
] (tA) {%
  {\ttfamily
  \$ ls\\
  \$ find\\
  \$ grep -rn\\
  \$ sed -n\\
  \$ cat}
};

\node[
  card,
  anchor=north,
  align=center
] (agA) at ($(tA.south)+(0,-0.62)$) {%
  \textbf{Model under evaluation}
};

\node[
  card,
  anchor=north,
  below=0.28cm of agA,
  fill=figvuln!6,
  draw=figvuln!45,
  align=center,
  text width=6.14cm,
  inner xsep=2pt
] (subA) {%
  {\scriptsize\ttfamily
  submit\_vulnerable\_files(["proxy/proxy.go"])}
};

\node[
  card,
  anchor=north,
  below=0.28cm of subA,
  align=center
] (scoreA) {%
  \textbf{File F1}\\[-1pt]
  {\scriptsize patch-touched files as labels}
};

% ============================================================
% Phase B
% ============================================================

\node[
  repo,
  anchor=north,
  below=0.22cm of hB
] (snapB) {%
  \textbf{\textcolor{figclean!45!black}{Patched snapshot}}
  \hfill post-fix\\[4pt]
  {\ttfamily
  zalando/skipper/\\
  \hspace*{1.5ex}proxy/\\
  \hspace*{4ex}\textcolor{figclean}{\textbf{proxy.go}}\\
  \hspace*{1.5ex}\ldots}
};

\node[
  note,
  anchor=north west,
  font=\scriptsize\itshape
] (roB) at ($(snapB.south west)+(0,-0.12)$)
  {read-only terminal};

\node[
  term,
  anchor=north,
  below=0.48cm of snapB
] (tB) {%
  {\ttfamily
  \$ ls\\
  \$ find\\
  \$ grep -rn\\
  \$ sed -n\\
  \$ cat}
};

\node[
  card,
  anchor=north,
  align=center
] (agB) at ($(tB.south)+(0,-0.62)$) {%
  \textbf{Model under evaluation}
};

\node[
  card,
  anchor=north,
  below=0.28cm of agB,
  fill=figclean!6,
  draw=figclean!45,
  align=center
] (subB) {%
  {\scriptsize\ttfamily
  submit\_no\_vulnerability\_found()}
};

\node[
  card,
  anchor=north,
  below=0.28cm of subB,
  align=center
] (scoreB) {%
  \textbf{True Negative Rate}\\[-1pt]
  {\scriptsize correctly reports no vulnerability}
};

% ============================================================
% Internal flow
% ============================================================

\draw[arr] (snapA) -- (tA);
\draw[arr] (snapB) -- (tB);

\draw[arr] (tA.south) -- (agA.north);
\draw[arr] (tB.south) -- (agB.north);

% Tool-call labels moved to the side of the arrow
\node[
  note,
  anchor=west
] at ($($(tA.south)!0.5!(agA.north)$)+(0.18,0)$)
  {$\leq 15$ tool calls};

\node[
  note,
  anchor=west
] at ($($(tB.south)!0.5!(agB.north)$)+(0.18,0)$)
  {$\leq 15$ tool calls};

\draw[arr] (agA) -- (subA);
\draw[arr] (agB) -- (subB);

\draw[arr] (subA) -- (scoreA);
\draw[arr] (subB) -- (scoreB);

% ============================================================
% Sandbox boundaries
% ============================================================

\begin{pgfonlayer}{background}

  \node[
    draw=black!45,
    dashed,
    rounded corners=3pt,
    inner xsep=5pt,
    inner ysep=5pt,
    fit=(snapA)(roA)(tA)(agA)
  ] (sbA) {};

  \node[
    draw=black!45,
    dashed,
    rounded corners=3pt,
    inner xsep=5pt,
    inner ysep=5pt,
    fit=(snapB)(roB)(tB)(agB)
  ] (sbB) {};

\end{pgfonlayer}

% ============================================================
% Legend
% ============================================================

\begin{pgfonlayer}{background}
  \node[
    fit=(scoreA)(scoreB),
    inner sep=0pt
  ] (scoregroup) {};
\end{pgfonlayer}

\node[
  rounded corners=2pt,
  draw=black!30,
  fill=black!2,
  anchor=north,
  text width=12.5cm,
  inner sep=5pt,
  align=center,
  font=\scriptsize
] at ($(scoregroup.south)+(0,-0.32)$) {%
  \tikz[baseline=-0.5ex]{
    \draw[
      dashed,
      black!45,
      line width=0.7pt
    ] (0,0) -- (0.55,0);
  }
  \quad
  \textbf{Sandbox:}
  read-only repository
  \quad$\cdot$\quad
  network disabled
  \quad$\cdot$\quad
  no writes
  \quad$\cdot$\quad
  no code execution
};

\end{tikzpicture}

\caption{
\textbf{Task structure.}
Each task is constructed from a GitHub Security Advisory by pairing its weakness
class with vulnerable and patched snapshots of the same repository. The model
receives only the generic CWE description and read-only repository access;
advisory text, CVE identifiers, affected versions, and patch-derived file labels
are withheld. Phase~A asks the model to localize vulnerable files and is scored
against files touched by the security patch. Phase~B presents the patched
snapshot and measures whether the model correctly reports no vulnerability.
Both phases use the same system prompt, tools, and interaction budget; only the
repository state changes.
}
\label{fig:task}
\end{figure}

Current evaluations capture localization only indirectly. Existing security benchmarks largely fall into two settings: those that preselect the code to be analyzed, and those that evaluate a downstream security outcome. In the former, models judge functions or fragments that have already been retrieved, making the task one of vulnerability recognition rather than repository search~\cite{ding2025primevul}. In the latter, agents may search a full codebase, but success is defined by whether they repair or reproduce vulnerable behavior; the relevant location may therefore be supplied explicitly~\cite{bui2026vul4py} or remain unobserved because correctness is determined through execution~\cite{shi2026cybergyme2e}. In either case, localization is not the object of evaluation. This leaves a basic capability poorly characterized: given a weakness to investigate and an unfamiliar repository, can an agent identify the files in which that weakness is implemented?

We introduce the \textbf{Vulnerability Localization Benchmark (VLoc Bench)} to evaluate this capability directly. We construct VLoc Bench from public GitHub Security Advisories (GHSAs) \cite{github2026advisorydb}, which link disclosed vulnerabilities to affected repositories and their security fixes. Each of its 500 tasks pairs a repository snapshot from immediately before the corresponding security fix with the snapshot after it, spanning 290 repositories across six packaging ecosystems. An agent receives the pre-fix repository together with the generic MITRE description of one weakness class~\cite{mitre2026cwe}, explores the codebase through a read-only terminal, and returns the implementation files it judges to contain the vulnerability. The weakness description specifies what type of vulnerability to search for, but provides no repository-specific evidence: the agent receives no advisory text, CVE identifier, fixing commit, file hint, or line range. We score the returned set against the implementation files modified by the security patch using file-level $F_1$. A second phase presents the corresponding patched repository under the same interface, where the recorded vulnerability has been removed and the correct response is to report no file. We then use this benchmark to study how localization varies with model scale, task-specific training, repository structure, and agent search behavior.

Our results show that repository-scale vulnerability localization remains difficult for current systems. The strongest configuration only reaches 0.229 File~$F_1$. Difficulty also depends strongly on the repository being searched: localization performance falls substantially as repositories become larger and relevant code becomes more dispersed. At the same time, successful localization does not necessarily correlate with correctly recognizing when the vulnerability has been removed. Systems that perform well in Phase~A can still report vulnerable files in already-patched repositories, revealing a tension between aggressively searching for vulnerabilities and avoiding unsupported reports. Evaluating localization alone can favor systems that report broadly, whereas evaluating only patched repositories can favor systems that rarely commit to a localization. VLoc Bench therefore evaluates both sides of the problem: whether an agent can identify security-relevant implementation when a vulnerability is present, and whether it can refrain from reporting that vulnerability after remediation.

\section{Related Work}
\label{sec:related_work}

Vulnerability localization connects two lines of work that have developed largely in parallel. Software-engineering research studies how agents navigate repositories and identify code relevant to a reported issue, while cybersecurity benchmarks increasingly place agents in real codebases to detect, reproduce, or repair vulnerabilities. These literatures share repository-scale reasoning as a common challenge, but differ in what they make observable: software-engineering benchmarks increasingly score where an agent looks, whereas security benchmarks have primarily scored what an agent concludes or does after reasoning about vulnerable code. Table~\ref{tab:related} summarizes the individual benchmarks; here, we focus on the broader themes that connect them.

\paragraph{Repository-Scale Localization in Software Engineering}

Localization is an established problem in software engineering: given a reported defect, identify the files, functions, or regions of a repository relevant to resolving it. Classical work formulated this as ranking source files against bug reports, with files modified by the eventual developer fix serving as ground truth; recent agentic work extends the same formulation to systems that actively search repositories and choose their own retrieval sets~\cite{zhou2012buglocator,ye2014learningtorank,li2026contextbench,chen2026codegrep,jiang2025cosil,zhang2026sweexplore,zhang2026corebench,chen2025locagent}. This shift makes precision as important as recall: an agent must not only recover relevant files, but avoid filling its context with unrelated code~\cite{li2026contextbench,chen2026codegrep}. The task nevertheless begins from a known software issue. An issue report establishes that something is wrong and often supplies repository-specific evidence about the failure; indeed, recent benchmarks explicitly filter reports that reveal the answer location because such information can make retrieval nearly solved before repository search begins~\cite{zhang2026corebench}. Thus, software engineering provides a mature formulation of repository localization, but typically under the assumption that a concrete defect has already been reported.

\begin{table}[!t]
\centering
\footnotesize
\setlength{\tabcolsep}{3pt}
\hyphenpenalty=10000

\caption{Related work by scored output, what the prompt discloses, and whether localization and behavior on patched code are scored. Instance counts follow each paper's reported unit.}
\label{tab:related}

\begin{tabular}{@{}>{\raggedright\arraybackslash}p{2.9cm}>{\raggedright\arraybackslash}p{1.7cm}@{\hskip 6pt}r@{\hskip 8pt}>{\raggedright\arraybackslash}p{1.9cm}>{\raggedright\arraybackslash}p{2.2cm}>{\raggedright\arraybackslash}p{2.1cm}>{\raggedright\arraybackslash}p{2.1cm}@{}}
\toprule
& Scored output & Inst. & Scope & Location disclosed & Localization & Patched code \\
\midrule

\multicolumn{7}{@{}l}{\emph{Repair and completion}} \\
Vul4Py \cite{bui2026vul4py} & Patch & 100 & Python & Files, line ranges & --- & --- \\
SecRepoBench \cite{shen2026secrepobench} & Completion & 318 & C/C++ & Region masked out & --- & --- \\

\midrule
\multicolumn{7}{@{}l}{\emph{Triggering}} \\
CyberGym \cite{wang2026cybergym} & Crashing input & 1,507 & C/C++ & Approximate, in description & --- & Pre/post criterion \\
BountyBench \cite{zhang2025bountybench} & Exploit & 40 & 25 codebases & Four conditions & --- & Detect snapshots \\

\midrule
\multicolumn{7}{@{}l}{\emph{Detection}} \\
PrimeVul \cite{ding2025primevul}\textsuperscript{a}
& Label & 6,968 fn. & C/C++ & Fragment supplied & --- & 5,480 pairs \\

JitVul \cite{yildiz2025jitvul}\textsuperscript{b}
& Label & 879 fn. & C/C++ & Commit supplied & --- & 879 pairs \\

RepoPairBench (DREA) \cite{sun2026drea}
& Label & 200 & Python & Function, file path & --- & False-positive rate \\

VulEval \cite{wen2024vuleval}\textsuperscript{c}
& Label, ranked deps. & 232,239 fn. & C/C++ & Function supplied & Dependencies, Pre@$k$ & --- \\

SecVulEval \cite{ahmed2025secvuleval}\textsuperscript{c}
& Label, statements & 25,440 fn. & C/C++ & Function supplied & Statements, 300 judged & Fixed versions included \\

VulDetectBench \cite{liu2024vuldetectbench}
& Label; lines & 1,000; 100 & C/C++ & One vuln., under 4K tokens & Lines, LoC recall & --- \\

\midrule
\multicolumn{7}{@{}l}{\emph{Localization, general software issues}} \\
ContextBench \cite{li2026contextbench} & Context set & 1,136 & 8 languages & Issue report & Files, F1 & --- \\
CodeGrep \cite{chen2026codegrep} & File list & Undisclosed & Python & Issue report & Files, $F_{\beta=0.5}$ & --- \\
CoSIL \cite{jiang2025cosil} & Ranked files, functions & 800 & Python & Issue report & Files, Top-$k$ & --- \\
SWE-Explore \cite{zhang2026sweexplore} & Ranked regions & 848 & 10 languages & Issue report & Files, recall & --- \\
CORE-Bench \cite{zhang2026corebench} & Ranked chunks & 5,061 & 11 languages & Query, leaks filtered & Chunks, NDCG@$k$ & --- \\
Loc-Bench (LocAgent) \cite{chen2025locagent} & Ranked locations & 560 & Python & Issue report & Files, Acc@$k$ & --- \\

\midrule
\multicolumn{7}{@{}l}{\emph{Localization, real vulnerabilities}} \\
RustMizan \cite{elsayed2026rustmizan} & Locations & 173 & Rust & None; 96 reduced & Functions, lines, F1 & 78, not split out \\
CWE-Bench-Java (IRIS) \cite{li2025iris} & Dataflow paths & 120 & Java & Weakness class & Methods, one-sided & --- \\
VulnGym \cite{ji2026vulngym} & Entry, operation, trace & 408 & 23 repositories & Advisory text & Files, oracle only & --- \\
AutoTrace \cite{zibaeirad2026autotrace} & Trigger statements & 744 & C/C++ & Fixing commit & Statements, hit rate & 771 pairs, SinkTrace \\

\midrule
VLoc Bench\textsuperscript{d}
& Files & 500 & 6 ecosystems & Weakness class & Files, F1 & True negative rate \\

\bottomrule
\end{tabular}

\vspace{3pt}
\begin{minipage}{\linewidth}
\scriptsize
\textsuperscript{a} PrimeVul's instance count refers to vulnerable functions rather than all functions in the corpus. \\
\textsuperscript{b} JitVul's instance count refers to vulnerable functions; its matched pairs are drawn from PrimeVul, so the two rows are not independent corpora. \\
\textsuperscript{c} VulEval and SecVulEval report counts over all functions in their respective corpora rather than only vulnerable functions. \\
\textsuperscript{d} To our knowledge, no prior benchmark scores file-level vulnerability localization over a whole repository with the location withheld while also evaluating the corresponding patched repository.
\end{minipage}

\end{table}

\paragraph{Localization Is Implicit in Security Evaluation}

Security evaluation has traditionally exposed a different part of the analysis pipeline. Detection benchmarks ask whether code is vulnerable after a function, snippet, commit, or other candidate region has already been selected; this measures vulnerability recognition, not the preceding problem of finding that region in a repository~\cite{ding2025primevul,yildiz2025jitvul,wen2024vuleval,ahmed2025secvuleval,liu2024vuldetectbench}. Repair benchmarks similarly focus on whether vulnerable behavior can be removed and may therefore provide localization information explicitly, while reproduction and exploitation benchmarks can evaluate success through execution without consulting a source-code location at all~\cite{bui2026vul4py,shen2026secrepobench,wang2026cybergym,lee2025secbench}. As security evaluation has moved toward complete repositories, however, the search required before these downstream actions has become increasingly visible. Performance changes sharply with the amount of vulnerability-specific information supplied---from a weakness class or report to crash traces and fixing patches---and end-to-end evaluations identify vulnerability discovery as a major bottleneck when such evidence is withheld~\cite{wang2026cybergym,shi2026cybergyme2e,zhang2025bountybench}. Repository search is therefore already part of security-agent performance, even when it is not itself the scored output.

\paragraph{Vulnerability Localization as Repository Search}

Recent work has begun to bring these two perspectives together by scoring locations associated with real vulnerabilities~\cite{elsayed2026rustmizan,li2025iris,ji2026vulngym,zibaeirad2026autotrace}. These evaluations make clear that ``localizing a vulnerability'' can refer to several related targets: vulnerable functions or statements, security-relevant data-flow paths, entry and critical-operation points, or statements that trigger vulnerable behavior. They also vary in how much of the search problem remains for the system: the analysis may begin from an entire repository, a reduced file or function, a vulnerability-specific advisory, a weakness specification, or the fixing commit itself~\cite{elsayed2026rustmizan,li2025iris,ji2026vulngym,zibaeirad2026autotrace,li2024interpvd}. These formulations address different stages of vulnerability analysis, but together they establish an important point: security reasoning and repository localization can be evaluated separately from the downstream act of producing an exploit or patch. This motivates treating vulnerability localization as a security-aware repository-search problem, where the system must connect an abstract weakness to the concrete implementation that realizes it.

\paragraph{Localization When the Vulnerability May Be Absent}

A second distinction emerges when localization is considered in a security setting. Conventional bug localization assumes that the issue described in the prompt exists, so failing to return a candidate location is simply a localization failure~\cite{jiang2025cosil,chen2025locagent,zhang2026corebench}. Security tools, by contrast, must also operate on code in which a suspected vulnerability is absent or has already been remediated. False positives have therefore long been treated as a practical concern in static analysis, and recent security evaluations increasingly use paired vulnerable and patched code or pre/post-patch execution to measure whether systems continue to report behavior after a fix~\cite{bessey2010billion,sun2026drea,wang2026cybergym,zhang2025bountybench,ding2025primevul,yildiz2025jitvul,zibaeirad2026autotrace}. Bringing this distinction to repository-scale localization changes the decision being evaluated: a system must determine not only \emph{where} evidence for a weakness lies, but whether the repository supports reporting a location at all. Localization and restraint therefore become complementary aspects of the same security-analysis problem and the basis for our work in this paper.

\section{Benchmark Design}
\label{sec:benchmark_design}

VLoc Bench evaluates whether an agent can identify vulnerable files in a repository given only a weakness class. The design rests on one principle: since security matters in any codebase, the task stays repository-agnostic. Nothing specific to the repository enters it, so every file the agent names comes from its own search. Each task links a weakness class to two repository states and a set of vulnerable implementation files derived from the fixing patch. The remainder of this section describes how these states and labels are constructed, how agents interact with them, and how their predictions are scored. We use CWE descriptions as the task specification because they identify the class of security weakness to search for without exposing the repository-specific localization cues often present in advisories or vulnerability reports. More broadly, CWE provides a common taxonomy through which the same localization task can be defined across otherwise heterogeneous repositories, languages, and software ecosystems.

\subsection{Benchmark Construction}
\label{subsec:benchmark_construction}

\paragraph{Threat model.} Vulnerability localization can target different points in the lifecycle of a weakness. Prior work has considered the \emph{trigger site}, where vulnerable state manifests as an unsafe operation~\cite{zibaeirad2026autotrace,li2024effectivenessfunctionlevelvulnerabilitydetectors}, while patch-based localization instead identifies the implementation that must be modified to remediate the vulnerability. We adopt the latter, defense-oriented view: given that a vulnerability of a particular CWE may exist somewhere in a repository, the agent must identify the files a defender would need to investigate and remediate. Accordingly, VLoc Bench defines the localization target as the \emph{patch site} and uses the implementation files modified by the eventual security fix as ground truth. Whereas many cybersecurity evaluations are organized around the capability being measured---detecting, triggering, exploiting, or repairing a vulnerability---our formulation also makes explicit who the localization signal is intended to serve, defenders. Beyond matching the needs of triage and remediation, this definition gives us a reproducible target that can be recovered from real-world vulnerability fixes, enabling the benchmark construction described next.

\paragraph{Sourcing.} Each task begins with a GitHub Security Advisory (GHSA)~\cite{github2026advisorydb}, a record linking a weakness class (CWE)~\cite{mitre2026cwe} to a repository and the commit that resolved it. The fixing commit provides the repository states and the patch used to derive the vulnerable-file labels.

\paragraph{Repository states.}  Each task comprises two snapshots of the same repository, one preceding and one following the security fix. The \textit{pre-push} state is the commit immediately before the patch and contains the vulnerability. The \textit{post-push} state is the patch commit itself and contains the fixed implementation. Both snapshots preserve the directory structure, configuration files, and source code. The paired states support localization before the fix and verification after it.

\paragraph{Ground-truth labels.} We derive the vulnerable-file labels deterministically from the patch diff. The label set contains every implementation file modified by the patch, excluding files that are exclusively test files. We detect test and documentation files using path-based heuristics, including files under \texttt{test/}, \texttt{tests/}, \texttt{spec/}, \texttt{\_\_tests\_\_/}, \texttt{doc/}, and \texttt{docs/} directories or matching \texttt{*\_test.*}, \texttt{*\_spec.*}, \texttt{test\_*.*}, \texttt{*.md}, and \texttt{*.rst} patterns. Thus, the released label sets contain only implementation source files changed by the security fix; they exclude tests, documentation, and configuration files. Using the implementation files changed by a fix as ground truth follows established repository-scale localization practice, including recent agent benchmarks \cite{zhou2012buglocator,ye2014learningtorank,chen2025locagent,li2026contextbench}.

\paragraph{Quality control.} We exclude tasks where: (1) the patch modifies only tests, documentation, CI configuration, or lock files; (2) the repository has been deleted or made private since the advisory; (3) the patch spans more than 50 files; or (4) the advisory lacks a CWE classification. The final benchmark contains 500 tasks from advisories disclosed between 2016 and 2026, including 21 disclosed in 2026.

\paragraph{Benchmark composition.} The 500 tasks span 6 package ecosystems: Go (n=215, 43\%), Maven (n=104, 21\%), npm (n=88, 18\%), pip (n=52, 10\%), Rust (n=40, 8\%), and Composer (n=1, $<$1\%). They cover 147 unique CWE categories, with the most frequent being CWE-400 (Resource Exhaustion, n=53), CWE-20 (Improper Input Validation, n=45), CWE-200 (Information Exposure, n=27), CWE-22 (Path Traversal, n=27), and CWE-770 (Allocation without Limits, n=23). By CVSS severity: Critical (n=57, 11\%), High (n=219, 44\%), Medium (n=194, 39\%), and Low (n=30, 6\%). 78\% of tasks carry assigned CVE identifiers. Repository sizes range from 12~KB to 840~MB (median 4.2~MB), and ground-truth file counts have a median of 3, with 28.4\% of tasks containing a single ground-truth file and 14.8\% containing ten or more.

\subsection{Evaluation Protocol}
\label{subsec:evaluation_protocol}

\paragraph{Task inputs and outputs.} Each task is evaluated in two phases using the paired repository snapshots. The model receives only the CWE category description (e.g., ``CWE-79: Improper Neutralization of Input During Web Page Generation'') and terminal access to the repository. We withhold advisory text, file hints, severity scores, affected version ranges, and CVE identifiers from the model's prompt. In \textbf{Phase~A} (localization), the model receives the pre-push repository and submits the files that contain the vulnerability. In \textbf{Phase~B} (verification), it receives the post-push repository and the same CWE description, then declares that no vulnerability is present.

Both phases use the same system prompt, tools, submission options, and command budget; only the repository state changes. In Phase~A, the model must submit one or more vulnerable file paths to receive a nonzero File~F1. Calling \texttt{submit\_no\_vulnerability\_found} or exhausting the budget without submitting a prediction is incorrect. In Phase~B, the correct outcome is that no vulnerable file path is submitted. Calling \texttt{submit\_no\_vulnerability\_found}, exhausting the budget without a submission, or returning no prediction is scored as a true negative; submitting any file path is a false positive. We use this shared interface for all evaluations to ensure consistent comparisons across models and phases.

\paragraph{Agent interface.} All models use the same system prompt and agent interface. The terminal exposes read-only commands for inspecting the repository, including file listing, text search, file viewing, and metadata queries. Write operations, network access, and process execution are prohibited. The model terminates a task by calling \texttt{submit\_vulnerable\_files} with a ranked list of paths or \texttt{submit\_no\_vulnerability\_found}. It has a budget of 15 terminal commands and 20 generation turns; three consecutive turns without a tool call trigger forced termination. These constraints are identical across all evaluated systems. We use temperature 0.3 with a 16{,}384-token completion limit; runs are stochastic, and we report the mean across three independent runs.

\paragraph{Sandbox.} Each task runs in a fresh Docker container (Ubuntu 24.04) with 2 CPU cores, 4~GB RAM, network disabled, and a 10-second per-command timeout. The container is destroyed after each task, ensuring no information leakage between evaluations. The repository is mounted read-only at \texttt{/repo/}. Together, these restrictions isolate each evaluation and prevent external access, repository modification, and cross-task information leakage.

\subsection{Scoring}
\label{subsec:scoring}

VLoc Bench scores the model's final submission under the same interface in both phases. The model searches a repository snapshot and then either submits vulnerable file paths or declares that no vulnerability is present. The score depends on the repository state: Phase~A rewards overlap with patch-derived files, while Phase~B rewards a correct clean-repository judgment. The phases therefore require different metrics. Phase~A is a set-retrieval task: its ground-truth set may contain several implementation files, and the score must penalize both omissions and unrelated predictions. Phase~B has an empty ground-truth set by construction, so it requires a binary measure of whether the model declares the patched repository clean. We use File~F1 for Phase~A and TNR for Phase~B, and report abstain rate separately.

\paragraph{File F1 (Phase~A).} We score localization with File~F1, the harmonic mean of file-level precision and recall:
\begin{align}
    \text{Precision} &= \frac{|\text{submitted} \cap \text{ground\_truth}|}{|\text{submitted}|}, \quad
    \text{Recall} = \frac{|\text{submitted} \cap \text{ground\_truth}|}{|\text{ground\_truth}|} \\
    \text{File F1} &= \frac{2 \cdot \text{Precision} \cdot \text{Recall}}{\text{Precision} + \text{Recall}}
\end{align}
If the model submits no files or calls \texttt{submit\_no\_vulnerability\_found} on a vulnerable repository, File~F1 is 0. The submitted paths are treated as an unordered set, so path order does not affect the score. This follows recent repository-localization benchmarks that evaluate final file predictions with file-level precision--recall measures~\cite{li2026contextbench,chen2026codegrep,sutawika2026codescout}.

\paragraph{Metric integrity.} Prior work aligns evaluation with the observable outcome of the task. Patch-generation benchmarks such as SWE-bench assess whether a submitted change satisfies the repository's behavioral checks, while work on agentic cybersecurity capabilities commonly uses execution-based criteria to test whether a generated proof of concept reproduces the target behavior across pre- and post-patch builds~\cite{jimenez2024swebench,wang2026cybergym}. Other agent benchmarks verify state-changing actions against task-specific oracle annotations. These criteria are appropriate when correctness is expressed through program behavior or environment state. VLoc Bench instead asks the model to submit a final set of repository paths, so Phase~A is evaluated against patch-derived file labels with File~F1. The precision--recall formulation distinguishes incomplete localization from overinclusive predictions, which an exact-match or recall-only score would not. The same principle applies to Phase~B: its binary clean-repository output is scored with TNR. Across both phases, the score is determined solely by the final submission, independent of tests, execution traces, and output formatting.

\paragraph{True Negative Rate (Phase~B).} We score verification with the true negative rate (TNR), the fraction of Phase~B tasks where the model correctly calls \texttt{submit\_no\_vulnerability\_found}:
\begin{equation}
    \text{TNR} = \frac{|\text{tasks correctly declared clean}|}{|\text{total Phase~B tasks}|}
\end{equation}
Submitting any file path on a patched repository is a false positive and does not count as a true negative.

\paragraph{Abstain Rate.} We also report the fraction of Phase~A tasks where the model fails to submit a localization prediction, either by calling \texttt{submit\_no\_vulnerability\_found} on a vulnerable repository or by exhausting its turn budget. An explicit clean-code judgment or an unfinished run counts as an abstention. Submitting incorrect files does not: it is a localization attempt and receives File~F1 of 0 if none of the submitted paths is correct.

\paragraph{Macro-aggregation.} For each run, we compute File~F1 separately for each task and report the macro-average over the 500 tasks. TNR and abstain rate are computed as proportions over their respective phase tasks, which is equivalent to averaging binary task-level outcomes. We then report the mean of each run-level metric across three independent runs. File~F1 is therefore a macro-averaged task-level metric rather than a pooled file-level statistic.

Together, File~F1 and TNR measure the two required decisions: which files to report in the pre-push state and whether to report no vulnerability in the post-push state. Existing benchmarks such as CyberGym and ExploitGym assess agentic cybersecurity capabilities that can directly support offensive workflows, including vulnerability reproduction and exploitation~\cite{wang2026cybergym,wang2026exploitgym}. VLoc Bench instead targets a more neutral, defensive capability: locating the implementation files associated with a weakness in source code. Agents use a read-only interface and cannot modify or execute the repository, keeping the evaluation focused on analysis rather than operational exploitation. We view exploitation and vulnerability localization as distinct capabilities, and design the benchmark construction to preserve that distinction.

\section{Experiments}
\label{sec:experiments}

\subsection{Models Evaluated}
\label{subsec:models_evaluated}

We evaluate \nummodels{} models and four static-analysis tools. The model suite spans frontier closed-source systems, open-weight models across the $\left[350\mathrm{M},\,753\mathrm{B}\right]$ parameter range, and models specialized for code search or vulnerability localization. This coverage allows us to compare general-purpose capability with parameter scale, domain-specific training, and non-agentic baselines. The evaluated systems are:

\begin{itemize}
    \item \textbf{Frontier (closed-source):} GPT-5.5 with default and xhigh reasoning effort~\cite{openai2026gpt55}, GPT-5, GPT-5 Mini, and GPT-5 Nano~\cite{singh2026openaigpt5card}; Gemini 3 Pro~\cite{google2026gemini3pro}, Gemini 2.5 Flash~\cite{geminiteam2025gemini25}, and Gemini 3.1 Flash-Lite~\cite{google2026gemini31flashlite}.
    \item \textbf{Open-weight large ($\geq$20B):} GLM-5.2 (753B)~\cite{glmteam2026glm5}, MiniMax-M2.7 (229B)~\cite{chen2026minimaxm2}, Qwen3.5-122B, Qwen3.5-35B-A3B, and Qwen3.5-27B~\cite{qwenteam2025qwen3}, GPT-OSS-120B and GPT-OSS-20B~\cite{openai2025gptoss}, Llama-3.3-70B~\cite{llamateam2024llama3}, and Gemma-4-31B~\cite{gemmateam2026gemma4}.
    \item \textbf{Open-weight small ($<$20B) and specialized:} CodeScout-14B~\cite{sutawika2026codescout}, Qwen3.5-9B~\cite{qwenteam2025qwen3}, Gemma-4-E4B (4B) and Gemma-4-E2B (2B)~\cite{gemmateam2026gemma4}, Antares-3B, Antares-1B, and Antares-350M~\cite{vijay2026antares}, and Granite-4.0-Micro (3B), Granite-4.0-1B, and Granite-4.0-350M~\cite{graniteteam2025granite4}.
    \item \textbf{Static analysis tools (non-model baselines):} Semgrep and Semgrep-CWE~\cite{semgrep2025}, CodeQL~\cite{github2025codeql}, and Horusec~\cite{zupit2025horusec}.
\end{itemize}

We keep the task interface fixed across model evaluations. Open-weight models are served with vLLM~\cite{kwon2023vllm} on H100 GPUs, while closed-source models use their respective API endpoints. Detailed serving and generation settings are provided in Appendix~\ref{sec:eval_cost}. Static-analysis tools run directly on each repository with their default rulesets; Semgrep-CWE uses CWE-specific rules matched to each task's classification. The system prompt and tool schemas are included in the released code repository. All reported scores are means over three independent evaluation runs.

\subsection{Main Results}
\label{subsec:main_results}

Table~\ref{tab:main_leaderboard} summarizes Phase~A performance across all evaluated systems. We compare model families and static-analysis baselines before examining how localization quality varies with parameter scale and domain-specific training; Phase~B verification is reported separately below.

\begin{table}[!htbp]
\centering
\caption{Phase~A performance on VLoc Bench (File F1, Precision, Recall). Systems are grouped by category and sorted by File~F1 within each group. The five highest-scoring model systems achieve at least 0.18 File~F1, while all remaining systems score at most 0.16.}
\label{tab:main_leaderboard}
\renewcommand{\arraystretch}{1.15}
\setlength{\tabcolsep}{6pt}
\begin{tabular}{lrcccc}
\toprule
\textbf{Model} & \textbf{Params} & \textbf{Open} & \textbf{File F1} & \textbf{Precision} & \textbf{Recall} \\
\midrule
\multicolumn{6}{c}{\textit{Frontier (Closed-Source)}} \\
\midrule
GPT-5.5 (xhigh) & --- & \texttimes & 0.229 & 0.310 & 0.221 \\
GPT-5.5 (default) & --- & \texttimes & 0.221 & 0.305 & 0.211 \\
Gemini 3 Pro & --- & \texttimes & 0.152 & 0.190 & 0.153 \\
Gemini 2.5 Flash & --- & \texttimes & 0.102 & 0.132 & 0.098 \\
GPT-5 Mini & --- & \texttimes & 0.098 & 0.115 & 0.096 \\
Gemini 3.1 Flash Lite & --- & \texttimes & 0.095 & 0.131 & 0.090 \\
GPT-5 & --- & \texttimes & 0.048 & 0.062 & 0.048 \\
GPT-5 Nano & --- & \texttimes & 0.024 & 0.038 & 0.021 \\
\midrule
\multicolumn{6}{c}{\textit{Open-Weight ($\geq$20B)}} \\
\midrule
GLM-5.2 & 753B & \checkmark & 0.186 & 0.226 & 0.186 \\
Gemma-4-31B & 31B & \checkmark & 0.101 & 0.131 & 0.097 \\
Qwen3.5-27B & 27B & \checkmark & 0.091 & 0.116 & 0.088 \\
Qwen3.5-122B & 122B & \checkmark & 0.091 & 0.124 & 0.083 \\
Qwen3.5-35B-A3B & 35B & \checkmark & 0.085 & 0.115 & 0.081 \\
GPT-OSS-20B & 20B & \checkmark & 0.070 & 0.095 & 0.065 \\
GPT-OSS-120B & 120B & \checkmark & 0.069 & 0.095 & 0.062 \\
MiniMax-M2.7 & 229B & \checkmark & 0.054 & 0.078 & 0.050 \\
Llama-3.3-70B & 70B & \checkmark & 0.012 & 0.016 & 0.014 \\
\midrule
\multicolumn{6}{c}{\textit{Open-Weight ($<$20B)}} \\
\midrule
CodeScout-14B & 14B & \checkmark & 0.044 & 0.065 & 0.039 \\
Qwen3.5-9B & 9B & \checkmark & 0.043 & 0.058 & 0.039 \\
Gemma-4-E2B & 2B & \checkmark & 0.039 & 0.045 & 0.042 \\
Gemma-4-E4B & 4B & \checkmark & 0.034 & 0.039 & 0.034 \\
Granite-4.0-350M & 350M & \checkmark & 0.001 & 0.001 & 0.001 \\
Granite-4.0-Micro & 3B & \checkmark & 0.000 & 0.000 & 0.000 \\
Granite-4.0-1B & 1B & \checkmark & 0.000 & 0.000 & 0.000 \\
\midrule
\multicolumn{6}{c}{\textit{Specialized Models}} \\
\midrule
Antares-3B & 3B & \checkmark & 0.223 & 0.298 & 0.219 \\
Antares-1B & 1B & \checkmark & 0.209 & 0.268 & 0.221 \\
Antares-350M & 350M & \checkmark & 0.135 & 0.144 & 0.176 \\
\midrule
\multicolumn{6}{c}{\textit{Static Analysis Tools}} \\
\midrule
Semgrep & N/A & \checkmark & 0.086 & 0.091 & 0.155 \\
Semgrep-CWE & N/A & \checkmark & 0.052 & 0.057 & 0.071 \\
CodeQL & N/A & \checkmark & 0.023 & 0.025 & 0.030 \\
Horusec & N/A & \checkmark & 0.020 & 0.021 & 0.038 \\
\bottomrule
\end{tabular}
\end{table}

The leaderboard shows substantial variation across systems, with no single category uniformly dominating. Because the table brings together general-purpose models, domain-specialized systems, and static-analysis baselines, the overall ordering should be read as a comparison of different approaches to repository search rather than as a single capability ranking. We therefore examine how model scale, training, and system design relate to localization performance in the analyses that follow.

Parameter count does not provide a reliable ordering among general-purpose models. Across open-weight and closed-source families, larger systems often underperform smaller counterparts, and the aggregate relationship between scale and File~F1 is only moderate. The GPT family makes this especially visible: GPT-5.5 substantially outperforms GPT-5, while GPT-5 Mini also exceeds GPT-5 despite its smaller size. % CITE NEEDED (unverified): the cyber-use safety claim below has no source. Wikipedia's GPT-5.5 article references https://openai.com/index/gpt-5-5-with-trusted-access-for-cyber/ (7 May 2026), which looks like the right page, but openai.com returns 403 to automated fetches so this was NOT read. Open it in a browser before citing it.
GPT-5.5's release explicitly considered cyber-use safety, suggesting that differences in model behavior and safety alignment across generations contribute to the ordering beyond parameter count alone. GPT-5's system card reports greater conservatism on dual-use cybersecurity tasks, including refusals in agentic security evaluations, while GPT-5 Mini performs better on Cyber Range evaluations~\cite{singh2026openaigpt5card}.

More generally, general software-engineering capability is not a reliable proxy for cybersecurity performance: a system may underperform because safety alignment constrains work in a dual-use setting, or because security-specific reasoning is out of distribution relative to its software-engineering and coding training.

Static-analysis tools remain reliable, widely available open-source baselines: they outperform several language models, although their fixed rules provide limited coverage when the relevant implementation files are not known in advance. Their closest resource-efficient LLM counterparts, models below 10B parameters that can run on consumer hardware, do not achieve comparable localization performance without task-specific training. Thus, interactive search alone is insufficient at this scale: static-analysis tools remain competitive and are displaced only by models fine-tuned for vulnerability localization.

\subsection{False Positive Verification Results}
\label{subsec:phase_b}

\begin{table}[!htbp]
\centering
\caption{Phase~B verification results for all 27 evaluated models, reported as true negative rate (TNR). TNR is the fraction of patched repositories correctly identified as clean; lower values indicate more false-positive vulnerability reports.}
\label{tab:phase_b}
\renewcommand{\arraystretch}{1.15}
\setlength{\tabcolsep}{6pt}
\begin{tabular}{lrc}
\toprule
\textbf{Model} & \textbf{Params} & \textbf{TNR} \\
\midrule
\multicolumn{3}{c}{\textit{Frontier (Closed-Source)}} \\
\midrule
GPT-5 Nano & --- & 0.868 \\
GPT-5 & --- & 0.743 \\
GPT-5 Mini & --- & 0.702 \\
Gemini 3.1 Flash Lite & --- & 0.632 \\
Gemini 2.5 Flash & --- & 0.392 \\
Gemini 3 Pro & --- & 0.329 \\
GPT-5.5 (xhigh) & --- & 0.279 \\
GPT-5.5 (default) & --- & 0.192 \\
\midrule
\multicolumn{3}{c}{\textit{Open-Weight ($\geq$20B)}} \\
\midrule
Qwen3.5-122B & 122B & 0.750 \\
Qwen3.5-27B & 27B & 0.748 \\
Llama-3.3-70B & 70B & 0.745 \\
Qwen3.5-35B-A3B & 35B & 0.740 \\
GPT-OSS-120B & 120B & 0.720 \\
GPT-OSS-20B & 20B & 0.719 \\
Gemma-4-31B & 31B & 0.682 \\
GLM-5.2 & 753B & 0.582 \\
MiniMax-M2.7 & 229B & 0.321 \\
\midrule
\multicolumn{3}{c}{\textit{Open-Weight ($<$20B) and Specialized}} \\
\midrule
Granite-4.0-1B & 1B & 1.000 \\
Granite-4.0-350M & 350M & 0.976 \\
Granite-4.0-Micro & 3B & 0.863 \\
Gemma-4-E4B & 4B & 0.845 \\
Qwen3.5-9B & 9B & 0.814 \\
Gemma-4-E2B & 2B & 0.755 \\
CodeScout-14B & 14B & 0.563 \\
\midrule
\multicolumn{3}{c}{\textit{Specialized Models}} \\
\midrule
Antares-3B & 3B & 0.034 \\
Antares-1B & 1B & 0.008 \\
Antares-350M & 350M & 0.012 \\
\midrule
\multicolumn{3}{c}{\textit{Static Analysis Tools}} \\
\midrule
Semgrep & N/A & 0.912 \\
Semgrep-CWE & N/A & 0.996 \\
CodeQL & N/A & 0.988 \\
Horusec & N/A & 0.980 \\
\bottomrule
\end{tabular}
\end{table}

Table~\ref{tab:phase_b} summarizes Phase~B results for all 27 evaluated models. The ranking differs substantially from Phase~A, showing that successful localization does not by itself imply reliable verification. This contrast is visible even within the frontier systems: GPT-5.5 (xhigh), which leads Phase~A localization, is comparatively prone to reporting files in patched repositories, whereas GPT-5 Nano shows the opposite profile, with weak localization but strong restraint. The Granite base models make the abstention caveat clear: high TNR can result from declining to engage with the task rather than from recognizing that no vulnerability is present. TNR therefore captures a distinct behavioral requirement, calibrated restraint when the repository is clean. Taken together, the two phases expose the difference between pursuing evidence of a weakness and withholding an unsupported claim, making their joint evaluation necessary for distinguishing useful security analysis from indiscriminate alerting.

These results show that current models capable of vulnerability localization are not uniformly calibrated against false positives. Because alert fatigue is a persistent concern in cybersecurity operations, improving localization without preserving the ability to recognize clean code is insufficient for practical use. As we build more capable agentic systems for cybersecurity, false-positive control should remain a first-class requirement for genuinely usable defensive assistance.

\subsection{Scaling Analysis}
\label{subsec:scaling}

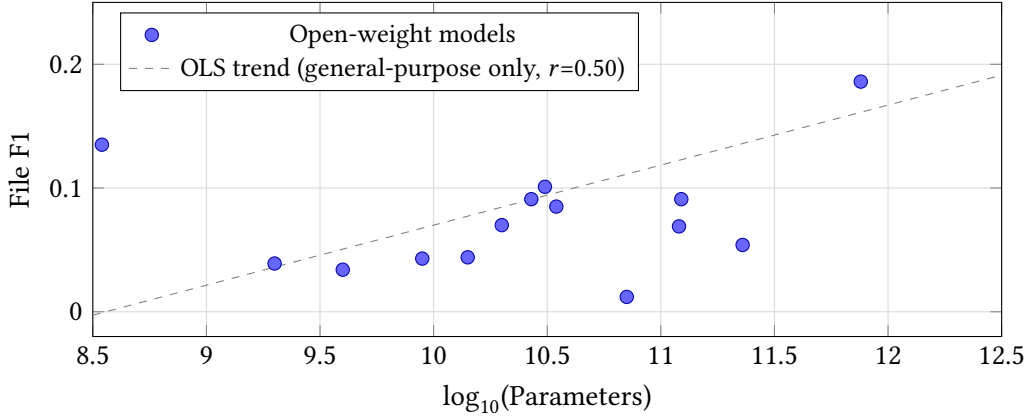
\begin{figure}[!htbp]
\centering
\begin{tikzpicture}
\begin{axis}[
  width=0.85\linewidth,
  height=6cm,
  xlabel={$\log_{10}$(Parameters)},
  ylabel={File F1},
  xmin=8.5, xmax=12.5,
  ymin=-0.02, ymax=0.25,
  grid=major,
  grid style={gray!30},
  mark size=2.5pt,
  legend style={at={(0.03,0.97)}, anchor=north west, font=\small},
]
\addplot[only marks, mark=*, color=blue!70!black, mark options={fill=blue!60}]
  coordinates {
    (8.54,0.135) (9.0,0.209) (9.48,0.223)
    (9.30,0.039) (9.60,0.034) (9.95,0.043) (10.15,0.044)
    (10.43,0.091) (10.49,0.101) (10.54,0.085)
    (10.85,0.012) (11.09,0.091) (10.30,0.070)
    (11.36,0.054) (11.08,0.069) (11.88,0.186)
  };
\addlegendentry{Open-weight models}

\addplot[dashed, gray, domain=8.5:12.5, samples=2] {0.0485*x - 0.415};
\addlegendentry{OLS trend (general-purpose only, $r$=0.50)}
\end{axis}
\end{tikzpicture}
\caption{File~F1 as a function of $\log_{10}$(parameter count) for open-weight models. The ordinary least squares (OLS) trend is fitted only to the 13 general-purpose models with available parameter counts; domain-specialized models are shown separately. Mixture-of-experts models are positioned by total parameter count rather than active parameter count.}
\label{fig:scaling}
\end{figure}

To summarize the association between model size and localization performance, we fit an ordinary least squares (OLS) regression, which estimates the linear trend that minimizes the squared differences between observed and predicted File~F1 values. The fit uses the 13 general-purpose open-weight models with available parameter counts. Figure~\ref{fig:scaling} examines the relationship between File~F1 and parameter count for open-weight systems. The fitted relationship is positive but accompanied by substantial scatter: model size provides some signal, yet does not determine localization quality. Models with comparable sizes can perform quite differently, and smaller systems can exceed much larger ones, yielding a non-monotonic scaling pattern. This pattern also appears in software-engineering localization. CodeScout~\cite{sutawika2026codescout} reports that smaller models trained directly on file-level localization rewards can outperform much larger base models, while SWE-Bench Pro shows similar reversals across general-purpose systems~\cite{deng2025swebenchpro}. These comparisons motivate asking whether the same scaling behavior holds when localization is conditioned on a security weakness.

Software-engineering localization and cybersecurity localization are related but distinct capabilities. VLoc Bench requires an agent to connect a CWE's security semantics to concrete implementation patterns in an unfamiliar repository and then identify the files implicated by that reasoning. General coding ability may support repository navigation, but it does not guarantee the security reasoning needed to recognize and localize the relevant weakness, particularly under dual-use alignment. The benchmark therefore measures security-aware localization rather than general-purpose scaling alone.

\paragraph{Evaluation cost.} Evaluation cost varies substantially across systems. A full 500-task Phase~A sweep costs between \$0.60 for Antares-350M (approximately 11 minutes on a single H100) and \$141 for GPT-5.5 xhigh via the OpenAI API (approximately 5 hours), a 170$\times$ range. GLM-5.2 via OpenRouter provides an intermediate reference at \$12.50 and approximately 50 minutes. Local open-weight models up to 31B parameters, served with 16 parallel workers on a single H100, complete evaluation in under one hour, making continuous benchmarking practical within a standard CI pipeline stage. Per-system cost and runtime details are provided in Appendix~\ref{sec:eval_cost}.

\section{Analysis}
\label{sec:analysis}

Aggregate performance leaves the sources of difficulty unresolved. We therefore examine which repository properties make localization difficult, whether model characteristics explain performance beyond the repository itself, which search behaviors accompany successful localization, and how unsuccessful trials differ in their decisions and tool use. Together, these analyses connect the properties of the searched codebase to the behavior of the searching agent, providing a basis for interpreting File~F1 beyond the leaderboard.

\subsection{Repository Difficulty Regression}
\label{subsec:repo_regression}

To identify which properties of a repository make a task intrinsically difficult, we fit a Lasso regression using features available before any model is evaluated. Lasso is a regularized linear regression that adds an $\ell_1$ penalty to coefficient magnitudes, shrinking weak associations toward zero and yielding an interpretable set of predictors. The fitted regression predicts per-task File~F1 from repository structure and task metadata, rather than from agent actions or outputs. This isolates benchmark-intrinsic difficulty from the behavior of the evaluated systems.

For each of the 500 tasks, we extract 42 features from the vulnerable repository snapshot and its associated metadata. Thirty-four repository-structural features summarize the source tree, including file counts and lines-of-code (LOC) distributions, directory topology, project signals such as CI, Docker, tests, and READMEs, and compressed and uncompressed size. Eight additional features describe vulnerability metadata, including CVSS score, the number of ground-truth files, associated CWEs and CVEs, disclosure year, and CWE super-categories covering input validation, memory, and resource management. We add 10 one-hot indicators for ecosystem membership and severity, giving 52 variables in total. The number of ground-truth files is included as a task-complexity covariate: it indicates how many files must be found, not which files constitute the answer.

The regression uses approximately 40,000 model--task--run observations (27 models $\times$ up to 500 tasks $\times$ 3 runs), with per-task File~F1 as the outcome. We fit LassoCV with 5-fold cross-validation over 100 log-spaced alpha values and standardize all features to zero mean and unit variance before fitting. The $R^2$ values reported in Table~\ref{tab:repo_lasso} are the in-sample fit and the corresponding five-fold cross-validated fit of this repo-only regression.

\begin{table}[!htbp]
\centering
\begin{tabular}{lrr}
\toprule
\textbf{Feature} & \textbf{Coef.} & \textbf{Direction} \\
\midrule
\texttt{top5\_loc\_share} & +0.082 & easier \\
\texttt{repo\_size\_bucket} & +0.038 & easier \\
\texttt{log10\_zip\_size\_kb} & $-$0.037 & harder \\
\texttt{total\_source\_loc} & +0.025 & easier \\
\texttt{max\_file\_loc} & $-$0.023 & harder \\
\texttt{pct\_files\_depth\_1} & +0.018 & easier \\
\texttt{eco\_go} & $-$0.017 & harder \\
\texttt{has\_ci} & $-$0.016 & harder \\
\texttt{depth\_std} & +0.016 & easier \\
\texttt{pct\_large\_files} & $-$0.015 & harder \\
\bottomrule
\end{tabular}
\caption{Top 10 repo-level Lasso coefficients (52 features, $R^2$=0.188, CV=0.164). Positive coefficients indicate features that make localization easier; negative indicate harder.}
\label{tab:repo_lasso}
\end{table}

The repo-only regression explains $R^2$=0.188 of the observed File~F1 variation in-sample and $R^2$=0.164 under five-fold cross-validation, with 48 of 52 features retaining nonzero weight. Table~\ref{tab:repo_lasso} reports the strongest associations. Code concentration is the clearest signal: repositories whose lines are concentrated in a small number of large files are easier to localize, consistent with search strategies reaching relevant code in fewer commands. In contrast, larger compressed repositories, deeper directory structures, and Go projects are associated with lower File~F1. The ground-truth file count carries negligible weight ($-$0.006), suggesting that the number of files to find is less informative than the structure in which those files are embedded.

The positive coefficient for \texttt{repo\_size\_bucket} and negative coefficient for \texttt{log10\_zip\_size\_kb} are not interpreted as opposing size effects: their signs reflect multicollinearity between a continuous size measure and its bucketed counterpart. After controlling for continuous size, the bucket variable captures a residual nonlinear pattern in which mid-sized repositories tend to be easier than both extremes. Severity and CWE category contribute negligible total weight (sum $|\beta|$ = 0.010). The similar in-sample and cross-validated fits suggest limited overfitting. Bootstrap resampling over 100 iterations provides a second check on coefficient stability. The leading predictor is nonzero in every resample and has a 95\% confidence interval of [+0.072, +0.091].

Repository structure is therefore the strongest measured source of task difficulty, although it does not explain all of the variation. The hardest cases are shaped less by the number of labeled files than by how widely relevant code is distributed through the repository, motivating the combined analysis of repository and model factors below.

\begin{figure}[!htbp]
\centering
\begin{tikzpicture}
\begin{axis}[
  xbar,
  width=0.92\linewidth,
  height=7cm,
  enlarge x limits=false,
  clip=false,
  xlabel={Sum $|\beta|$ (grouped Lasso importance)},
  xlabel style={font=\small, yshift=-2pt},
  symbolic y coords={Severity, Model Family, Ecosystem, Vuln.\ Metadata, Model Scale, Repo Size, Model Category, Repo Structure},
  ytick=data,
  y tick label style={font=\small},
  xmin=0, xmax=0.33,
  xtick={0,0.05,0.1,0.15,0.2,0.25,0.3},
  scaled x ticks=false,
  xticklabel style={
    /pgf/number format/fixed,
    /pgf/number format/precision=2,
    font=\footnotesize,
  },
  bar width=10pt,
  enlarge y limits=0.1,
  xmajorgrids=true,
  grid style={gray!20},
  nodes near coords,
  nodes near coords style={
    font=\tiny,
    /pgf/number format/fixed,
    /pgf/number format/precision=3,
    anchor=west,
  },
]
\addplot[fill=blue!50, draw=blue!70] coordinates
  {(0.285,Repo Structure) (0.063,Model Category) (0.055,Repo Size) (0.043,Model Scale) (0.034,Vuln.\ Metadata) (0.031,Ecosystem) (0.030,Model Family) (0.010,Severity)};
\end{axis}
\end{tikzpicture}
\caption{Grouped Lasso importance from the combined regression (67 features, $R^2$=0.241, ${\sim}$40k samples). Repository structure carries 4.5$\times$ more summed coefficient weight than model category.}
\label{fig:grouped_importance}
\end{figure}
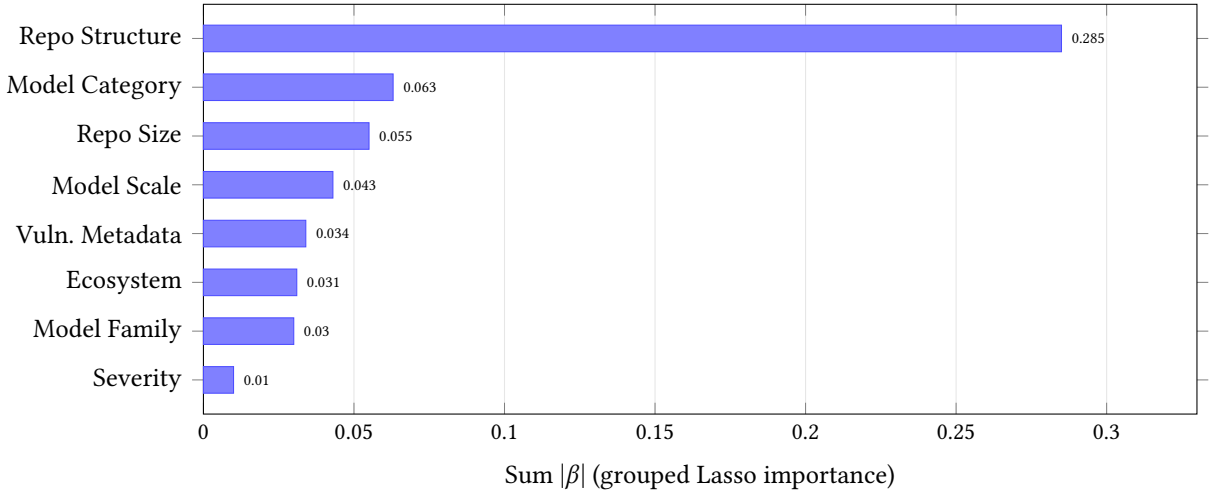

\subsection{Combined Regression: Structure Dominates}
\label{subsec:combined_regression}

To test whether model identity explains performance beyond repository structure, we combine 52 repository-level and 15 model-level features in a single Lasso regression. The combined regression uses 67 predictors, achieves $R^2$=0.241 in-sample and $R^2$=0.192 under cross-validation, and retains 57 nonzero coefficients. Figure~\ref{fig:grouped_importance} compares feature groups by their summed absolute coefficient magnitudes. Repository structure carries 4.5$\times$ more coefficient weight than model category (0.285 vs. 0.063).

To test whether this result depends on the Lasso penalty or on how feature-group importance is measured, we repeat the comparison using two alternatives. Elastic Net combines the Lasso's coefficient-shrinking penalty with an $\ell_2$ penalty; with an $l_1$-ratio of 0.5, the two penalties contribute equally. This model gives a 5.7$	imes$ ratio between repository structure and model category. Group-level permutation importance instead measures the reduction in $R^2$ when the values of one feature group are shuffled while the remaining features are left unchanged; this gives a 3.2$	imes$ ratio.

The ratios are also affected by how the features vary in the dataset. Repository features vary across 500 tasks, whereas model features repeat across only 27 systems, so the two groups do not contribute comparable amounts of independent variation. Permutation importance reduces the dependence on coefficient magnitude while preserving the same qualitative ordering. Taken together, the Lasso, Elastic Net, and permutation analyses consistently associate repository structure with more predictive information than model category, but these ratios describe predictive association rather than a causal contribution.

\paragraph{Model identity in isolation.} We next ask how much task performance can be predicted from model descriptors without using repository features. We fit a model-only Lasso with 15 predictors covering parameter count, model category, and model family. The regression explains $R^2$=0.053 of the observed variation in-sample and $R^2$=0.007 under cross-validation, as reported in Table~\ref{tab:model_lasso}. Thus, model descriptors explain little of the task-level variation. This result is compatible with the moderate model-aggregate correlation ($r$=0.50, Section~\ref{subsec:scaling}), which ranks 13 model-level means. The model-only regression asks a different question: whether those descriptors predict which particular tasks a system will solve. The aggregate correlation describes average ordering, whereas the cross-validated regression tests per-task explanatory power. Its near-zero cross-validated $R^2$ indicates that model descriptors shift average performance but have limited ability to predict task-specific outcomes. These coefficients should be read as associations, not causal effects. In this regression, the specialized-training indicator has the largest positive coefficient, followed by parameter count (Table~\ref{tab:model_lasso}).

\begin{table}[!htbp]
\centering
\begin{tabular}{lrr}
\toprule
\textbf{Feature} & \textbf{Coef.} & \textbf{Interpretation} \\
\midrule
\texttt{cat\_specialized} & +0.053 & domain training helps \\
\texttt{log10\_params} & +0.043 & scale helps \\
\texttt{fam\_llama} & $-$0.015 & underperforms at scale \\
\texttt{fam\_gemma} & +0.006 & slight edge \\
\texttt{cat\_open-small} & $-$0.005 & small generalists struggle \\
\texttt{cat\_granite} & $-$0.005 & untrained baselines fail \\
\bottomrule
\end{tabular}
\caption{Model-level Lasso coefficients (15 features, $R^2$=0.053, CV=0.007). Coefficients are descriptive effect sizes; the near-zero CV indicates model identity has negligible entry-level predictive power.}
\label{tab:model_lasso}
\end{table}

Taken together, these results support VLoc Bench's measurement premise: cybersecurity localization reflects both the complexity of the repository being searched and the model's cybersecurity understanding and capability. Repository structure accounts for more variation than model identity, while model identity still shifts average performance but provides little information about which individual tasks a system will solve. The benchmark's repositories therefore define the evaluation landscape, while models serve as comparative probes of cybersecurity capability within it. Holding this task distribution fixed while interchanging agent harnesses also makes VLoc Bench a system-level evaluation setting, supporting comparisons along both model-capability and system-design axes.

Repository and model summaries provide a useful first-order account of localization difficulty, but 76\% of the observed variation remains unexplained. This residual likely reflects task-specific properties that aggregate features cannot capture, including code idioms, naming conventions, framework-specific import patterns, and whether the vulnerable logic is reachable through textual search. We next examine these structural effects more directly by decomposing performance across repository conditions.

\subsection{Difficulty Decomposition}
\label{subsec:difficulty}

The regression identifies repository structure as a broad source of difficulty; we now examine how that difficulty is distributed across concrete repository conditions. We consider repository size, ecosystem, and the subset of tasks that no evaluated system solves.

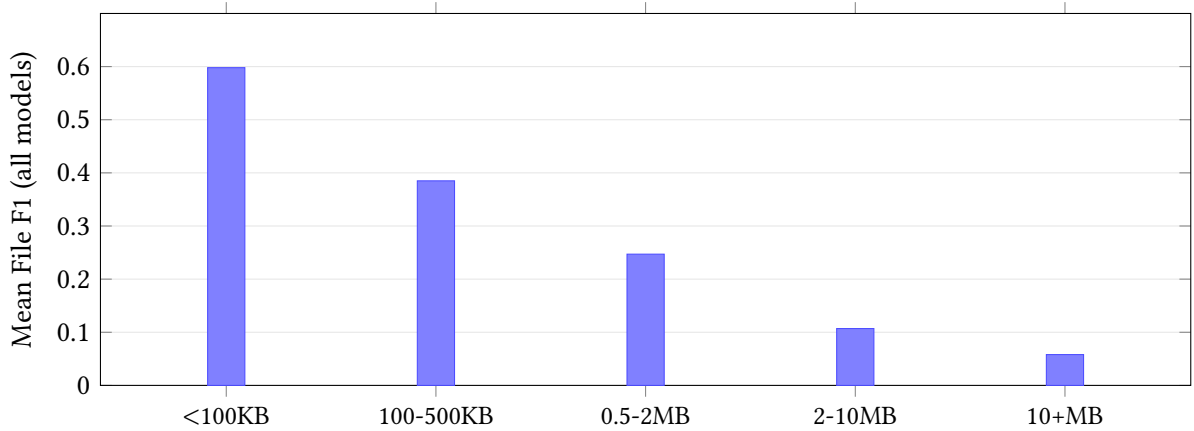
\begin{figure}[!htbp]
\centering
\begin{tikzpicture}
\begin{axis}[
  ybar,
  width=\linewidth,
  height=6.5cm,
  ylabel={Mean File F1 (all models)},
  symbolic x coords={$<$100KB, 100-500KB, 0.5-2MB, 2-10MB, 10+MB},
  xtick=data,
  x tick label style={font=\small},
  ymin=0, ymax=0.70,
  ytick={0, 0.10, 0.20, 0.30, 0.40, 0.50, 0.60},
  yticklabel style={/pgf/number format/fixed, /pgf/number format/precision=2},
  scaled y ticks=false,
  bar width=14pt,
  enlarge x limits=0.15,
  ymajorgrids=true,
  grid style={gray!20},
]
\addplot[fill=blue!50, draw=blue!70] coordinates
  {($<$100KB,0.598) (100-500KB,0.385) (0.5-2MB,0.247) (2-10MB,0.107) (10+MB,0.058)};
\end{axis}
\end{tikzpicture}
\caption{Mean File~F1 averaged across all \nummodels{} models, decomposed by repository size. Performance decays sharply: the smallest repositories ($<$100~KB, n=20) yield 10$\times$ higher scores than the largest (10+~MB, n=223). The majority of tasks (63\%) fall in the two hardest buckets ($\geq$2~MB).}
\label{fig:difficulty_by_size}
\end{figure}

\paragraph{Repository size.} Repository size provides the clearest broad separation in difficulty. Mean File~F1 falls from 0.598 for repositories smaller than 100~KB to 0.058 for repositories larger than 10~MB, a roughly 10$\times$ difference. The ordering is consistent across models: each model performs better on smaller repositories, while the relative ordering of models is mostly preserved across size buckets. This pattern is consistent with a larger search space reducing the signal available to search-based agents and making it harder for reasoning-based agents to retain the relevant code. It also parallels SWE-Bench Pro's finding that localization performance decreases as the number of ground-truth files grows~\cite{deng2025swebenchpro}. Repository scale therefore provides a strong, model-independent indicator of localization difficulty.

\paragraph{Ecosystem.} The size gradient does not, however, account for all of the observed differences. The ecosystem breakdown shows a clear ordering: Maven tasks are hardest on average, pip and npm tasks are easiest, and Go lies between these groups. Go is especially consequential for the benchmark because it accounts for 43\% of tasks, making its lower average performance a major component of the overall difficulty distribution. We interpret this variation as a consequence of ecosystem-specific coding conventions and project organization, which shape how relevant implementation is distributed across files and directories while remaining intertwined with overall repository size.

\paragraph{Unsolved tasks.} The interaction between repository scale and ecosystem is clearest among the tasks that remain unsolved. In 38.4\% of tasks, every model receives File~F1 = 0; large Go and Maven repositories account for 47\% and 31\% of this subset, respectively. Additional models or larger parameter counts do not remove this subset under the evaluated settings. The result is a persistent hard core of repository conditions that current localization systems do not reliably handle.

Taken together, these results show that difficulty is shaped by both the scale of the search space and the way code is organized within it. Repository size supplies the strongest general gradient, while the unsolved subset shows that current systems do not yet cover the full range of repository conditions. We next examine the search behavior behind these outcomes, asking whether successful localization is associated with deeper exploration, more targeted command use, or more reliable interaction with the tool interface.

\subsection{Behavioral Analysis}
\label{subsec:behavioral}

Aggregate scores describe whether a system succeeds, but not how it searches the repository. We therefore examine three observable properties of the recorded traces: exploration depth, command allocation, and tool reliability. These analyses operate on model-level behavior and identify associations with performance.

\paragraph{Exploration depth and performance.} We compute two summaries for each model from its Phase~A traces: the mean number of terminal commands issued per task-run and the mean File~F1 over those same task-runs. A Pearson correlation across the 27 paired model-level summaries gives $r$=0.72 ($p$<0.001). Models that use more of their available command budget therefore tend to achieve higher mean File~F1, with weaker systems more likely to terminate after limited exploration and stronger systems more likely to continue searching. This relationship should not be read as evidence that additional commands alone improve every task: among the strongest systems, command counts are already near the budget limit, so their remaining differences reflect the quality of the search rather than its duration. The pattern suggests that VLoc Bench rewards agentic, multi-turn, goal-directed tool use: each additional inspection can add substantial code to the context, making long-horizon search and context management central to effective localization.

\paragraph{Command strategy diversity.} Agents also differ in how they spend their commands. We assign every terminal operation in the Phase~A traces to one of three categories: search commands (e.g., \texttt{grep}, \texttt{rg}, and \texttt{find}), file-reading commands (e.g., \texttt{cat}, \texttt{head}, and \texttt{sed}), or exploratory commands (e.g., \texttt{ls} and \texttt{tree}). For each model, we divide the number of commands in each category by its total number of commands, producing a three-part command-use profile, and compare that profile with the model's mean File~F1. Higher-performing systems generally devote more of their interaction to targeted search and reading, while lower-performing systems spend more of their budget exploring the directory structure without a focused follow-up. Effective behavior is not tied to one fixed allocation: strong systems combine search and reading in different proportions, but they use exploration to narrow the search space rather than as an end in itself. This pattern is consistent with CodeScout's observation that trained agents converge toward targeted ripgrep-based search across different starting tool policies~\cite{sutawika2026codescout}.

\paragraph{Tool reliability.} The value of a command strategy also depends on whether the selected operations execute successfully. From each model's Phase~A traces, we compute a tool-error rate by dividing invalid commands, timeouts, and permission errors by the model's total issued commands, then compare this rate with its mean File~F1. Across the observed range, models with error rates in the low single digits tend to achieve higher localization scores, whereas rates above 15\% coincide with lower scores. Failed commands consume interaction budget without yielding repository evidence, which helps explain this association. Reliable tool use therefore appears necessary for effective search, but it is not sufficient: a model can execute commands cleanly and still fail to connect the CWE description to the relevant implementation. The behavioral evidence points to productive interaction, rather than command volume alone, as the more useful distinction.

Taken together, these traces show that higher localization performance is associated with sustained exploration, targeted search and reading, and reliable tool use. Command volume alone is not sufficient.

\subsection{Failure Mode Taxonomy}
\label{subsec:failure_taxonomy}

To complement File~F1, we classify every non-perfect Phase~A trial ($\text{File F1} < 1.0$) by the behavior recorded in its evaluation trace. Prior work such as SWE-Bench Pro uses model-based trajectory interpretation to assign semantic failure categories~\cite{deng2025swebenchpro}; VLoc instead uses command counts, precision, recall, and submission status. This summarizes how a trial ended without requiring an additional model to interpret the trajectory. The same trace produces the same label, which makes the classification reproducible across the ${\sim}$40,000 trials. Its limitation is that trace metadata does not identify semantic intent: it can distinguish early termination from an incorrect submission, but not whether an incorrect search reflected a misunderstood CWE or an unproductive choice of directory.

At Level~1, a failed trial is either \emph{Abstained}, meaning that the model submitted no files, or \emph{Submitted}, meaning that it submitted at least one file path but did not achieve perfect File~F1. Level~2 then separates these groups by observable behavior. Abstentions are \emph{Premature termination} when the model issues at most three terminal commands, \emph{Exhausted budget} when it issues at least 13 of the 15 allowed commands without submitting, and \emph{Inconclusive search} otherwise. A forced termination after three consecutive turns without a tool call follows the same command-count rule.

Submitted failures are divided by overlap with the ground-truth files and by submission breadth. \emph{Partial recall} denotes a submission with recall $>0$ and precision $\geq 0.5$, while \emph{Overly broad} denotes recall $>0$ and precision $<0.5$. Among submissions with no correct files, \emph{Wrong files} denotes submissions that are not overly broad, and \emph{Overly broad, zero recall} denotes submissions containing more than twice the ground-truth file count. These conditions make the Level~2 leaves mutually exclusive. Every non-perfect Phase~A trial therefore receives exactly one deterministic label.

\paragraph{Aggregate failure distribution.}
\label{subsubsec:aggregate_failures}

To characterize how localization attempts fail, we aggregate the deterministic labels across all non-perfect Phase~A trials. Table~\ref{tab:failure_aggregate} summarizes these outcomes across the 27 models and three runs per task, first separating trials that abstain from those that submit files and then dividing each group into more specific behavioral categories. Abstention accounts for the larger share of failures (59.1\%), while submitted failures account for the remainder (40.9\%). The two most common Level~2 categories are \emph{Exhausted budget} (32.3\%) and \emph{Wrong files} (27.4\%), covering trials that search without reaching a submission and trials that submit without identifying a correct file, respectively.

\begin{table}[!htbp]
\centering
\begin{tabular}{llr}
\toprule
\textbf{Level 1} & \textbf{Level 2} & \textbf{\%} \\
\midrule
\multirow{3}{*}{Abstained (59.1\%)} & Exhausted budget & 32.3 \\
 & Premature termination & 16.9 \\
 & Inconclusive search & 9.9 \\
\midrule
\multirow{4}{*}{Submitted (40.9\%)} & Wrong files & 27.4 \\
 & Partial recall & 8.5 \\
 & Overly broad & 2.6 \\
 & Overly broad (zero recall) & 2.4 \\
\bottomrule
\end{tabular}
\caption{Aggregate failure mode distribution across all 27 models (${\sim}$38,500 non-perfect trials, 3 runs per model--task pair). Exhausted budget and Wrong files together account for 59.7\% of failures.}
\label{tab:failure_aggregate}
\end{table}

Premature termination and partial recall represent different failure behaviors. Premature termination accounts for 16.9\% of failures and is associated with systems that stop after limited exploration, sometimes following malformed tool calls or difficulty parsing the initial repository structure. It is concentrated among smaller systems and uncommon among frontier systems. Partial recall accounts for 8.5\% of failures and is more common among stronger systems: these trials identify at least one relevant file but miss others, showing that multi-file localization remains difficult even when the search reaches the correct part of the repository.

\paragraph{Per-model failure profiles.}
\label{subsubsec:per_model_failures}

\begin{table*}[!htbp]
\centering
\renewcommand{\arraystretch}{1.18}
\setlength{\tabcolsep}{5.5pt}

\begin{tabular}{
l
@{\hskip 8pt}
r
@{\hskip 10pt}
r
@{\hskip 5pt}
r
@{\hskip 5pt}
r
@{\hskip 12pt}
r
@{\hskip 5pt}
r
@{\hskip 5pt}
r
@{\hskip 5pt}
r
}
\toprule
& &
\multicolumn{3}{c}{\textbf{Abstained (\%)}} &
\multicolumn{4}{c}{\textbf{Submitted (\%)}} \\
\cmidrule(lr){3-5}
\cmidrule(lr){6-9}

\textbf{Model}
& \textbf{F1\%}
& \textbf{Prem.}
& \textbf{Exh.}
& \textbf{Incon.}
& \textbf{Wrong}
& \textbf{Part.}
& \textbf{Broad}
& \textbf{Br.(0)} \\

\midrule

GPT-5.5        & 10.0 & 0.0  & 15.6 & 0.2  & 46.4 & 23.2 & 2.8  & 1.8 \\
Antares-3B     & 9.8  & 0.0  & 2.4  & 0.0  & 57.2 & 22.6 & 4.4  & 3.6 \\
Antares-1B     & 9.6  & 0.4  & 0.2  & 0.0  & 52.2 & 15.8 & 12.2 & 9.6 \\
GLM-5.2        & 8.6  & 0.0  & 28.6 & 0.8  & 38.4 & 16.2 & 5.0  & 2.4 \\
Gemini 3 Pro   & 5.3  & 0.0  & 26.9 & 0.2  & 41.6 & 15.2 & 6.7  & 4.2 \\
Qwen3.5-122B   & 5.4  & 0.0  & 66.2 & 4.0  & 16.2 & 7.4  & 0.6  & 0.2 \\
MiniMax-M2.7   & 2.4  & 0.2  & 28.0 & 0.2  & 59.5 & 6.0  & 0.6  & 3.0 \\
Llama-3.3-70B  & 0.6  & 75.3 & 0.0  & 0.8  & 19.8 & 1.2  & 0.4  & 1.8 \\
CodeScout-14B  & 2.7  & 5.1  & 1.8  & 52.0 & 34.3 & 3.9  & 0.0  & 0.2 \\
Antares-350M   & 3.8  & 0.8  & 0.0  & 0.2  & 53.6 & 7.4  & 17.0 & 17.2 \\

\bottomrule
\end{tabular}

\caption{
Failure mode profiles for 10 representative models (percentages of all 500 trials,
including successes). F1\% = perfect-score rate.
Prem.~= premature termination, Exh.~= exhausted budget,
Incon.~= inconclusive search, Wrong~= wrong files entirely,
Part.~= partial recall, Broad~= overly broad with some recall,
Br.(0)~= overly broad with zero recall.
Each model exhibits a distinct dominant failure mode.
}
\label{tab:failure_profiles}
\end{table*}

Table~\ref{tab:failure_profiles} shows that similar aggregate scores can arise from different failure distributions. Three recurring profiles are visible. \emph{Commit-heavy} systems rarely abstain but often submit incorrect files. \emph{Cautious exhausters} use most of their command budget without submitting, indicating that they search extensively but do not reach a decision within the available horizon. \emph{Premature terminators} stop after limited interaction, leaving little evidence that they engaged with the repository. These profiles summarize observable behavior rather than fixed model classes, and a single system may exhibit more than one across tasks.

Taken together, the failure profiles show that File~F1 can conceal materially different system behaviors. A model may stop after limited exploration, exhaust its budget without submitting, submit files with no overlap, or identify only part of a multi-file ground truth. These outcomes point to different limitations in tool use, search strategy, and submission decisions, but they are collapsed into the same aggregate score. The deterministic taxonomy preserves these distinctions across all trials, allowing future systems to be evaluated by whether they reduce premature termination, non-convergent search, incorrect submissions, and incomplete recall.

\section{Conclusion}
\label{sec:conclusion}

We introduce VLoc Bench to measure agentic vulnerability localization over complete software repositories, a capability that existing code and security benchmarks often leave implicit. Its paired repository states and agentic interface evaluate two complementary decisions: identifying the implementation files associated with a weakness when it is present, and recognizing that the repository is clean after the fix. File~F1 and TNR make both outcomes measurable under the same task conditions.

The results show that repository-scale localization remains far from solved. The strongest system achieves 0.229 File~F1, and 38.4\% of tasks receive no correct localization from any evaluated model. Beyond this aggregate difficulty, the analyses identify how performance varies across repositories and systems. Repository structure explains more variation than model descriptors, while parameter count alone provides only a moderate guide to performance. The behavioral and failure analyses further show that localization depends on sustained, targeted, and reliable tool use, and that strong localization does not guarantee restraint on patched code. These findings separate security-aware repository analysis from general software-engineering performance and from a model's willingness to submit an answer.

Most existing agentic cybersecurity benchmarks evaluate capabilities through CTF-style tasks, vulnerability reproduction, or exploit generation~\cite{shao2024nyuctf,wang2026cybergym,wang2026exploitgym}. These benchmarks measure important cybersecurity capabilities, but their success criteria center on triggering or exploiting a weakness rather than locating and triaging the affected implementation. VLoc Bench complements this literature by evaluating a more neutral defensive capability: source-code localization under a CWE-only prompt and read-only repository access, with Phase~B testing whether the agent avoids false positives after remediation. This positioning isolates a distinct stage of security work without requiring agents to modify, execute, or exploit the target code. We hope this work helps address this evaluation gap and supports both the agentic AI evaluation and cybersecurity communities.

\section{Limitations and Future Work}
\label{sec:limitations}

\paragraph{Limitations.}

VLoc Bench is scoped to the vulnerability associated with each advisory, rather than to the security of the repository as a whole. In Phase A, patch-touched implementation files provide a reproducible localization target, but may not capture every file relevant to the vulnerability. In Phase B, the patched snapshot establishes that the recorded vulnerability has been remediated, not that no other vulnerability exists. This limits evaluation of general-purpose static analyzers such as Semgrep, CodeQL, and Horusec: a finding outside the benchmark target may be valid but cannot be adjudicated without an additional oracle. Phase A is sensitive to both incomplete and overly broad localization through File F1, while Phase B provides only a binary target-vulnerability judgment. The benchmark remains reproducible through fixed repository snapshots, deterministic scoring, and repeated runs, although Phase B also conflates explicit abstention with failures such as turn exhaustion, API errors, or refusals when no submission is produced.

\paragraph{Future work.}

We plan to extend VLoc Bench by introducing levels of task context, from the current CWE-only setting to increasingly detailed vulnerability descriptions, to measure how additional security information changes localization performance. We also plan to strengthen Phase B by validating additional model findings through manual review, proofs of concept, or other behavioral oracles, allowing genuine discoveries to be separated from false positives. More broadly, vulnerability localization is only one stage of a security analyst's workflow. Future benchmarks should evaluate these intermediate capabilities intrinsically, alongside end-to-end extrinsic evaluations, so that progress can be measured not only by whether a system ultimately detects or repairs a vulnerability, but by which parts of the security-analysis pipeline it can reliably assist with.

\section{Ethics and Responsible Use}
\label{sec:ethics}

VLoc Bench is constructed
entirely from public GitHub Security Advisories describing vulnerabilities that are
already disclosed and already patched; we introduce no new vulnerability information
and release no exploits. The benchmark evaluates \emph{localization}---identifying
which files instantiate a known weakness class---not exploitation or weaponization,
and all agent interaction is confined to read-only terminal access within an isolated,
network-disabled sandbox. Our intent is to advance defensive tooling: measuring and
improving the ability of AI systems to help maintainers and security teams find
vulnerable code faster. We note the dual-use nature of any vulnerability-analysis
research and emphasize that automated localization should augment, not replace, expert
human review; as our results show, automated localization at repository scale remains
an open problem, and outputs from any current system should be treated as leads for
human review rather than conclusions. Because every entry corresponds to a fix that is
already public, the benchmark does not expand the attack surface of the included
projects beyond what is already documented in the public advisory record.

\bibliographystyle{plainnat}
\bibliography{bibliography}

\appendix
\section{Example Task}
\label{sec:example}

To make the task interface and scoring concrete, we show one representative Phase~A task together with selected steps from a successful and an unsuccessful run. As in the main evaluation, the models receive only the CWE description and read-only access to the pre-push repository.

\paragraph{Task T0aMD9EF.} The repository is \texttt{yahoo/serialize-javascript} (134~KB, npm ecosystem). The model receives:

\begin{quote}
\textit{CWE-79: Improper Neutralization of Input During Web Page Generation (`Cross-site Scripting') --- The product does not neutralize or incorrectly neutralizes user-controllable input before it is placed in output that is used as a web page that is served to other users.}
\end{quote}

The patch-derived label is \texttt{index.js}, the sole implementation file containing the affected serialization logic.

\paragraph{Successful run (GPT-5.5, xhigh).} The full run contains 15 terminal commands over 17 turns. The representative steps below show the progression from repository inspection to submission:

\begin{enumerate}
    \item lists the repository files with \texttt{find . -type f | sort};
    \item reads \texttt{index.js} in numbered segments using \texttt{nl} and \texttt{sed};
    \item searches for serialization and escaping logic with \texttt{rg}; and
    \item submits \texttt{[index.js]}, obtaining \textbf{File F1 = 1.0}.
\end{enumerate}

\paragraph{Unsuccessful run (Llama-3.3-70B).} The agent inspects the directory structure and issues several broad searches, but does not read \texttt{index.js} in sufficient depth to identify the affected logic. It exhausts the terminal budget without submitting a file list, resulting in \textbf{File F1 = 0.0}. This run illustrates an abstention after extended but inconclusive exploration.

The two runs differ in how they allocate the same interface and command budget. Even on a small repository with a single labeled implementation file, success depends on connecting the CWE description to the relevant code and committing the correct file list.

\section{Dataset Composition}
\label{sec:dataset_composition}
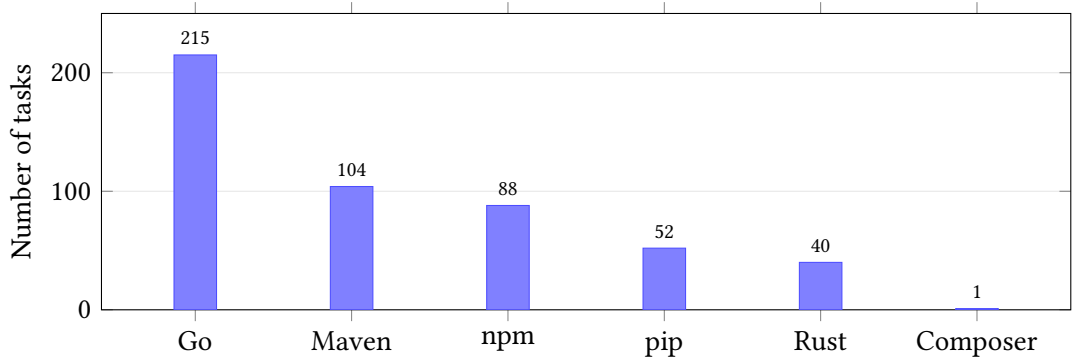
\begin{figure}[h]
\centering
\begin{tikzpicture}
\begin{axis}[
  ybar,
  width=0.9\linewidth,
  height=5.5cm,
  ylabel={Number of tasks},
  symbolic x coords={Go, Maven, npm, pip, Rust, Composer},
  xtick=data,
  ymin=0, ymax=250,
  bar width=16pt,
  enlarge x limits=0.12,
  ymajorgrids=true,
  grid style={gray!20},
  nodes near coords,
  every node near coord/.append style={font=\scriptsize},
]
\addplot[fill=blue!50, draw=blue!70] coordinates
  {(Go,215) (Maven,104) (npm,88) (pip,52) (Rust,40) (Composer,1)};
\end{axis}
\end{tikzpicture}
\caption{Distribution of VLoc Bench tasks across six package ecosystems. Go is the largest group (43\%), while Composer contributes one task.}
\label{fig:ecosystem_dist}
\end{figure}

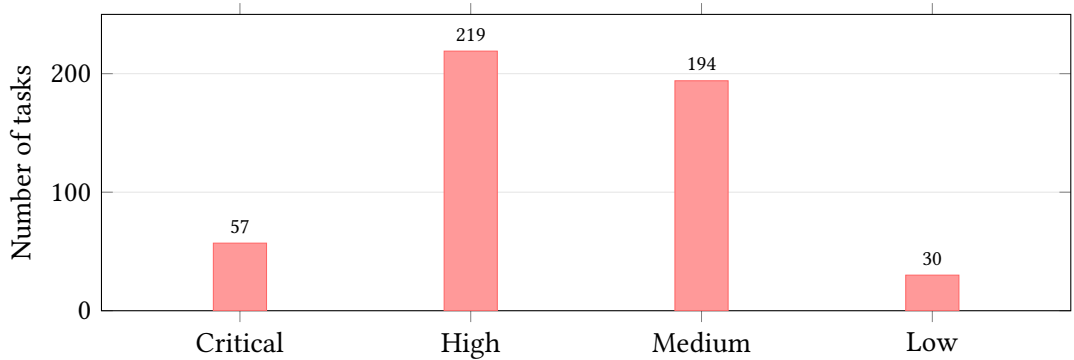
\begin{figure}[h]
\centering
\begin{tikzpicture}
\begin{axis}[
  ybar,
  width=0.9\linewidth,
  height=5.5cm,
  ylabel={Number of tasks},
  symbolic x coords={Critical, High, Medium, Low},
  xtick=data,
  ymin=0, ymax=250,
  bar width=20pt,
  enlarge x limits=0.2,
  ymajorgrids=true,
  grid style={gray!20},
  nodes near coords,
  every node near coord/.append style={font=\scriptsize},
]
\addplot[fill=red!40, draw=red!60] coordinates
  {(Critical,57) (High,219) (Medium,194) (Low,30)};
\end{axis}
\end{tikzpicture}
\caption{Distribution of VLoc Bench tasks by CVSS severity. High- and Medium-severity tasks together account for 83\% of the benchmark.}
\label{fig:severity_dist}
\end{figure}

VLoc Bench contains 500 tasks drawn from the natural composition of the GitHub Security Advisory database rather than from a deliberately balanced sampling scheme. We summarize the task distribution along two dimensions used in the main analysis: package ecosystem and CVSS severity. Ecosystem coverage is uneven: Go is the largest group (43\%), followed by Maven (21\%) and npm (18\%), while Composer contributes one task (Figure~\ref{fig:ecosystem_dist}). The severity distribution is concentrated in the High and Medium bands, which together account for 83\% of tasks (Figure~\ref{fig:severity_dist}). These distributions define the empirical support for the per-ecosystem and severity analyses in Section~\ref{sec:analysis}.

\section{Data and Code Availability}
\label{sec:availability}

The complete benchmark and the code needed to reproduce its evaluation are publicly available at \url{https://github.com/cisco-foundation-ai/vulnerability-localization-benchmark}. The release includes the manifest, snapshot retrieval and verification tooling, evaluation harness, scoring implementation, and analysis scripts:

\begin{itemize}
    \item \textbf{Data manifest} (\texttt{data/manifest.csv}): 500 tasks with repository metadata, commit SHAs for both vulnerable and patched states, CWE labels, ecosystem tags, and content-based MD5 checksums for reproducible verification.
    \item \textbf{Download and verification tooling}: scripts that retrieve repository snapshots as zip archives from GitHub at the exact commit SHAs recorded in the manifest, strip the archive prefix, resolve symlinks, and verify each task against its deterministic content MD5, computed over file paths and contents in lexicographic order.
    \item \textbf{Evaluation harness}: the full sandboxed evaluation pipeline including Docker image specification, agent protocol implementation, scoring code, and configuration files.
    \item \textbf{Scoring and analysis scripts}: code to compute File~F1, TNR, abstain rates, and to reproduce the statistical analyses reported in Section~\ref{sec:analysis}.
\end{itemize}

All 500 tasks are downloadable directly from public GitHub repositories at the recorded commit SHAs. The release does not require gated access or authentication.

\section{Evaluation Cost and Runtime}
\label{sec:eval_cost}

\paragraph{Model coverage.} Anthropic models (Claude Opus, Sonnet, and Haiku) are not included because their evaluation cost exceeds the available budget. A full 500-task sweep under the standardized harness would exceed \$600 at Claude Opus 4.8 pricing, while an unconstrained native-agentic configuration using Claude Code with subagent spawning costs \$1,658 for the same sweep. Future versions of VLoc Bench will incorporate these models as budget permits.

\paragraph{Serving configuration.} Open-weight models are served with vLLM (v0.19.1) on H100 GPUs using bfloat16 precision and a maximum sequence length of 32768. Native tool calling is enabled for Qwen3.5 and Gemma-4, and Antares uses a custom Granite chat template. Locally served models use temperature 0.3, a maximum of 16384 tokens per turn, and frequency penalty 0.3.

Closed-source models use their respective provider APIs with default temperature and penalty settings; GPT-5.5 xhigh additionally uses \texttt{reasoning\_effort=xhigh}. Because provider defaults are not fully documented, the system prompt, tool definitions, agent loop, and command budget are the controlled variables across model evaluations. Static-analysis tools run directly on each repository with their default rulesets, while Semgrep-CWE uses CWE-specific rules matched to each task's classification.

\begin{table}[h]
\centering
\caption{Wall-clock runtime and estimated cost for a full 500-task Phase~A evaluation. Local models are served via vLLM on a single H100 GPU with 16 parallel workers, with cost estimated from commodity H100 rental rates (\$2--4/hour). API-model costs use provider pricing at the time of evaluation (June 2026).}
\label{tab:cost}
\renewcommand{\arraystretch}{1.15}
\setlength{\tabcolsep}{6pt}
\begin{tabular}{lrrr}
\toprule
\textbf{System} & \textbf{Runtime} & \textbf{Total Cost} & \textbf{Cost/Task} \\
\midrule
Antares-3B (local, H100) & $\sim$15 min & \$0.82 & \$0.002 \\
Antares-1B (local, H100) & $\sim$13 min & \$0.71 & \$0.001 \\
Antares-350M (local, H100) & $\sim$11 min & \$0.60 & \$0.001 \\
GLM-5.2 (OpenRouter API) & $\sim$50 min & \$12.50 & \$0.025 \\
GPT-5.5 xhigh (OpenAI API) & $\sim$5 hrs & \$141.00 & \$0.282 \\
\bottomrule
\end{tabular}
\end{table}

\section{Lasso Regression Feature List}
\label{sec:feature_list}

Table~\ref{tab:full_features} enumerates the 67 features used in the combined Lasso regression described in Section~\ref{subsec:combined_regression}. The 52 repository-level features are extracted by static analysis of each vulnerable repository snapshot, while the 15 model-level features encode system identity and remain constant across tasks for a given system. All features are standardized to zero mean and unit variance before fitting.

\begin{table}[h]
\centering
\scriptsize
\begin{tabular}{p{2.4cm}p{5.8cm}r}
\toprule
\textbf{Group} & \textbf{Features} & \textbf{Count} \\
\midrule
Repo Structure & source\_file\_count, total\_source\_loc, avg\_file\_loc, median\_file\_loc, max\_file\_loc, top5\_loc\_share, pct\_large\_files, directory\_depth, avg\_directory\_depth, depth\_p90, depth\_std, n\_dirs\_depth\_3plus, pct\_files\_depth\_1, n\_directories, file\_ext\_entropy, max\_directory\_fanout, test\_to\_source\_file\_ratio, pct\_test\_loc, n\_languages, comment\_line\_ratio, has\_readme, has\_tests\_dir, has\_ci, has\_docker, n\_config\_files, n\_dependencies, n\_build\_scripts, n\_dotfiles & 28 \\
\midrule
Repo Size & log10\_repo\_size\_kb, log10\_zip\_size\_kb, log10\_merge\_size\_kb, compression\_ratio, size\_delta\_pct, repo\_size\_bucket & 6 \\
\midrule
Vuln.\ Metadata & cvss\_score, n\_gt\_files, n\_cwes, n\_cves, year, cwe\_input\_validation, cwe\_memory, cwe\_resource & 8 \\
\midrule
Ecosystem & eco\_go, eco\_npm, eco\_pip, eco\_maven, eco\_rust, eco\_composer & 6 \\
\midrule
Severity & sev\_critical, sev\_high, sev\_medium, sev\_low & 4 \\
\midrule
Model Scale & log10\_params & 1 \\
\midrule
Model Category & cat\_frontier, cat\_open-large, cat\_open-small, cat\_specialized, cat\_granite & 5 \\
\midrule
Model Family & fam\_gpt, fam\_gemini, fam\_qwen, fam\_gemma, fam\_gpt-oss, fam\_antares, fam\_granite, fam\_llama, fam\_other & 9 \\
\midrule
\textbf{Total} & & \textbf{67} \\
\bottomrule
\end{tabular}
\caption{Complete feature inventory for the combined Lasso regression. Repository-level features are extracted once per task; model-level features are constant across tasks for a given system.}
\label{tab:full_features}
\end{table}

The feature blocks preserve the distinction between task properties and system properties used throughout the analysis. Repository features vary across tasks, whereas model features repeat across the tasks evaluated for each system; this structure supports the repository-only, model-only, and combined specifications reported in Section~\ref{sec:analysis}.

\end{document}